\documentclass[sigconf, nonacm]{acmart}

\usepackage{graphicx}
\usepackage{array}
\usepackage[labelformat=simple]{subcaption}
\usepackage{amsmath}
\usepackage{wasysym }
\usepackage[export]{adjustbox}
\usepackage{url}
\usepackage{tikz}
\usepackage{balance}  
\usepackage{multirow}
\usepackage[normalem]{ulem}
\usepackage[linesnumbered,ruled,vlined]{algorithm2e}
\usepackage{enumitem}
\usepackage{makecell}

\newcommand*\circled[1]{\tikz[baseline=(char.base)]{
            \node[shape=circle,draw,inner sep=0.5pt] (char) {#1};}}

\newcommand{\cgsz}{\ensuremath{\mathop{cg\_size}}}

\SetKwProg{Fn}{Function}{ is}{end}
\SetAlFnt{\scriptsize}

\begin{document}
\title{Real-Time LSM-Trees for HTAP Workloads}

\author{Hemant Saxena}
\affiliation{%
  \institution{SAP Labs, Waterloo, Canada}
}
\email{h.saxena@sap.com}

\author{Lukasz Golab}
\affiliation{%
  \institution{University of Waterloo}
}
\email{lgolab@uwaterloo.ca}

\author{Stratos Idreos}
\affiliation{%
  \institution{Harvard University}
}
\email{stratos@seas.harvard.edu}

\author{Ihab F. Ilyas}
\affiliation{%
  \institution{University of Waterloo}
}
\email{ilyas@uwaterloo.ca}

\begin{abstract}
Real-time analytics systems employ hybrid data layouts in which data are stored in different formats throughout their lifecycle.  Recent data are stored in a row-oriented format to serve OLTP workloads and support high insert rates, while older data are transformed to a column-oriented format for OLAP access patterns.  We observe that a Log-Structured Merge (LSM) Tree is a natural fit for a lifecycle-aware storage engine due to its high write throughput and level-oriented structure, in which records propagate from one level to the next over time.  To build a lifecycle-aware storage engine using an LSM-Tree, we make a crucial modification to allow different data layouts in different levels, ranging from purely row-oriented to purely column-oriented, leading to a Real-Time LSM-Tree.  We give a cost model and an algorithm to design a Real-Time LSM-Tree that is suitable for a given workload, followed by an experimental evaluation of LASER - a prototype implementation of our idea built on top of the RocksDB key-value store. 
\end{abstract}

\maketitle



\section{Introduction} \label{sec:intro}
The need for real-time analytics or Hybrid Transactional-Analytical Processing (HTAP) is ubiquitous in applications such as content recommendation, real-time pricing, high-frequency trading, blockchains, and IoT \cite{htapSurvey}. These applications differ from traditional On-Line Transactional Processing (OLTP) and On-Line Analytical Processing (OLAP) applications in two aspects: 1) high data rates \cite{htapSurvey}; 2) access patterns change over the lifecycle of the data \cite{gartnerHTAP, facebook}. For example, recent data may be accessed via OLTP style operations (point queries and updates) as part of an alerting application, or to impute a missing attribute value based on the values of other attributes \cite{htapSurvey}. Additionally, recent and historical data may be accessed via OLAP style processes, perhaps to generate hourly, weekly and monthly reports, where all values of one column (or a few columns, depending on the time span of the report) are scanned \cite{htapSurvey}.

Traditionally, row-oriented layout was used for OLTP-heavy workloads and column-oriented layout for OLAP-heavy workloads. 
Recent systems,   
such as SAP HANA \cite{HANA}, SingleStore \cite{singlestore}, and IBM Wildfire \cite{wildfire} support real-time analytics 
using hybrid layouts, in which recent data are stored in a row-oriented format to serve point queries (OLTP), and older data are transformed to a column-oriented format suitable for OLAP.  Such systems can be described as having a \emph{lifecycle-aware} data layout.

We observe that a Log-Structured Merge (LSM) Tree is a natural fit for a lifecycle-aware storage engine.  LSM-Trees are used in  key-value stores (Google's BigTable and LevelDB, Cassandra, Facebook's  RocksDB), RDBMSs (Facebook's MyRocks, SQLite4), blockchains (e.g., Hyperledger uses LevelDB), and data stream and time series databases (e.g., InfluxDB).  
While Cassandra and RocksDB can simulate columnar storage via \emph{column families}, we are not aware of any lifecycle-aware key-value storage engines.
We fill this gap in our work, by designing a storage engine that can replace traditional LSM trees in the above applications 
to support real-time analytics.


An LSM-Tree is a multi-level data structure with a main-memory buffer and a number of secondary-storage levels with increasing size (details in Section~\ref{sec:background}). Periodically, or when full, the buffer is flushed to Level-0.  When Level-0, which stores multiple flushed buffers, is nearly full, its data are merged into the sorted runs residing in level one (via a \emph{compaction} process), and so on. We observe that LSM-Trees provide a natural framework for a lifecycle-aware storage engine for real-time analytics due to the following reasons. 
\\
\circled{1} \textbf{LSM-Trees are write optimized:} Writes and data transfers between levels are batched, allowing high write throughput.
\\
\circled{2} \textbf{LSM-Trees naturally propagate data through the levels over time:} At any point in time, the buffer stores the most recent data that have not yet been flushed (perhaps data inserted within the last hour), Level-0 may contain data between one hour and 24 hours old, and levels one and beyond store even older data.  
\\
\circled{3} \textbf{Different levels can store data in different layouts: }
Data may be stored in row format in the buffer and in some of the levels, and in column format in other levels.
This suggests a flexible and configurable storage engine that can be adapted to the workload.
\\
\circled{4} \textbf{Compaction can be used to change data layout:} 
Transforming the data from a row to a column format can be done during compaction, when a level is merged into the next level.




We make the following \textbf{contributions} in this paper.
\begin{itemize}[leftmargin=*]
\item 
We propose the \emph{Real-Time LSM-Tree}, which extends the traditional LSM-Tree with the ability to store data in a row-oriented or a column-oriented format in each level.
\item
We characterize the design space of possible Real-Time LSM-Trees.
To navigate this design space, we provide a cost model to select good designs for a given workload.
\item
We develop and evaluate LASER, a Lifecycle-Aware Storage Engine for Real-time analytics based on Real-Time LSM-Trees.  We implement LASER using RocksDB, which is a popular open-source key-value store based on LSM-Trees.  
\end{itemize}

\section{Overview of LSM-Trees} \label{sec:background}
\subsection{Design}
Compared to traditional read-optimized data structures such as B-trees, 
LSM-Trees focus on high write throughput while allowing indexed access to data \cite{chen}. LSM-Trees have two components: an in-memory piece that buffers inserts and a secondary storage piece. The in-memory piece consists of trees or skiplists, whereas the secondary storage piece consists of sorted runs.

\begin{figure}[t]
\centering
  \includegraphics[width=6.5cm,height=4cm]{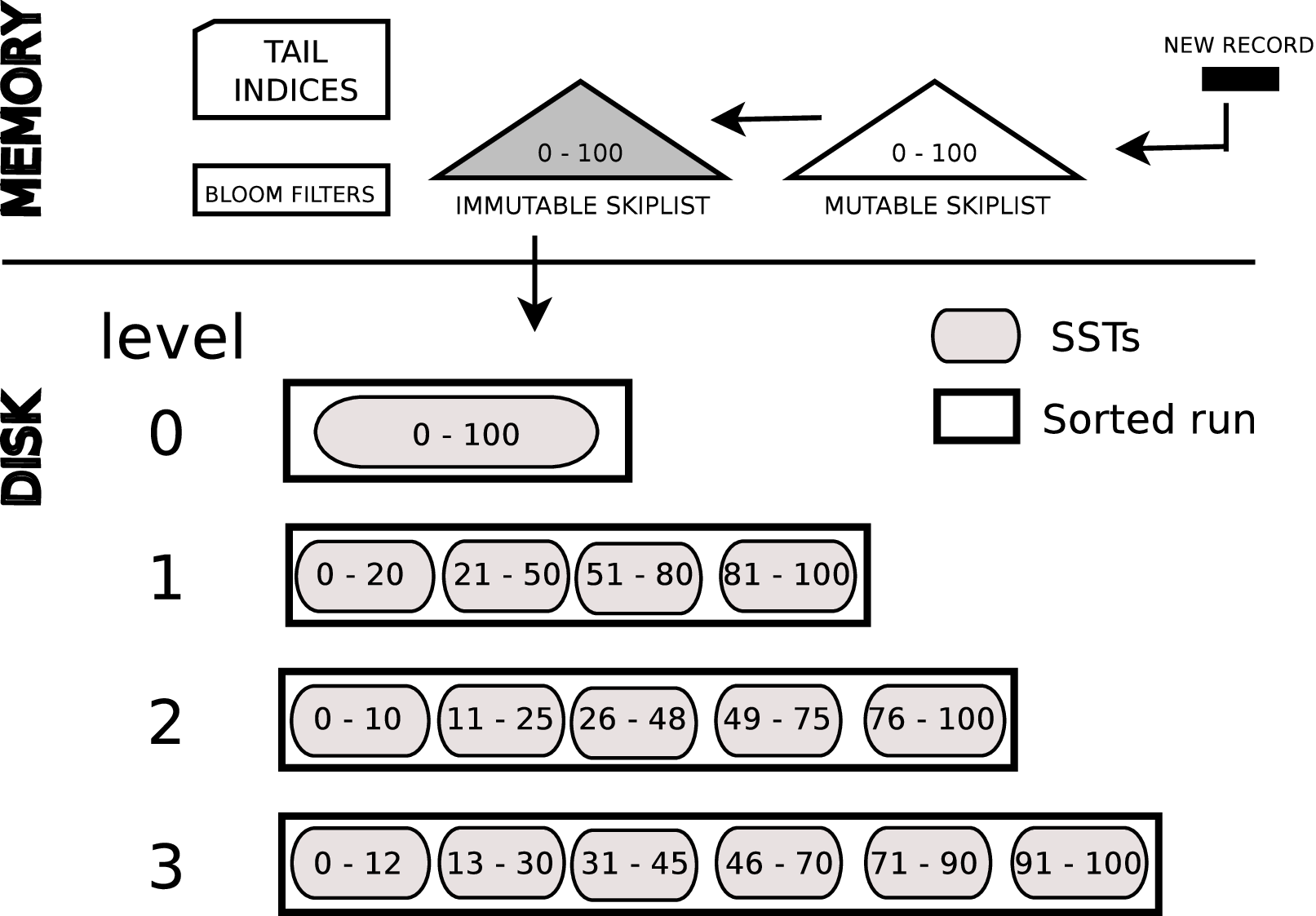}
  \caption{LSM-Tree with leveling merge strategy}
\label{fig:rocksdb}
\end{figure}

Figure~\ref{fig:rocksdb} shows the architecture of an LSM-Tree, with the memory piece at the top, followed by multiple levels of sorted runs on secondary storage (four levels, numbered zero to three, are shown in the figure). The memory piece contains two or more skiplists of user-configured size (two are shown in the figure).  New records are inserted into the most recent (mutable) skiplist and into a write-ahead-log for durability.  Once inserted, a record cannot be modified or deleted directly. Instead, a new version of it must be inserted and marked with a tombstone flag in case of deletions.


Once a skiplist is full, it becomes immutable and can be \emph{flushed} to secondary storage via a sequential write. Flushing is executed by a background thread (or can be called explicitly) and does not block new data from being inserted.
During flushing, each skiplist is sorted and serialized to a sorted run. Sorted runs are typically range-partitioned into smaller chunks called Sorted Sequence Tables (SSTs), which consist of fixed-size blocks. In Figure \ref{fig:rocksdb}, we show sorted runs being range-partitioned by key into multiple SSTs. For example, the sorted run in Level-1 has four SSTs; the first SST contains values for the keys in the range 0-20, the second in the range 21-50, and so on. Each SST contains a list of data blocks and an index block.  A data block stores key-value pairs ordered by key, and an index block stores the key ranges of the data blocks.

As sorted runs accumulate over time, query performance tends to degrade since multiple sorted runs may have to be accessed to find a record with a given key. To address this, sorted runs are merged by a background process called \textit{compaction}. The merging process organizes the disk piece into $L$ logical levels of increasing size with a size ratio of $T$.  For example, a size ratio of two means that every level is twice the size of the previous one.  In Figure \ref{fig:rocksdb}, we show four levels with increasing sizes.  The parameters $L$ and $T$ are user-configurable 
and their value depends on the expected number of entries in the database. 

Two common merging strategies are \textit{leveling} and \textit{tiering} \cite{dostoevsky,chen}. Their trade-offs are well understood: leveling has higher write amplification but is more read-optimized than tiering. Furthermore, the ``wacky continuum" \cite{wacky} provides tunable read/write performance by adjusting the merging strategy and size ratios. Our Real-Time LSM-Tree is independent of the merging strategy, but we use the leveling strategy in LASER since this is also used by RocksDB.

In leveling, each level consists of one sorted run, so the run at level $i$ is $T$ times larger than the run at level $i-1$. As a result, the run at level $i$ will be merged up to $T$ times with runs from level $i-1$ until it fills up. If multiple versions of the same key exist, then only the most recent version is kept, and any key with a tombstone flag is deleted. In practice, merging is done at SST granularity, i.e., some SSTs from level $i-1$ are merged with overlapping SSTs in level $i$. 
Sorted runs in Level-0 are not partitioned into SSTs (or have exactly one SST) because they are directly flushed from memory. Some implementations, such as RocksDB, make an exception for Level-0 and allow multiple sorted runs to absorb write bursts.

\begin{figure}[t]
\begin{subfigure}{0.23\textwidth}
  \centering
  \includegraphics[width=3.6cm,height=3.7cm]{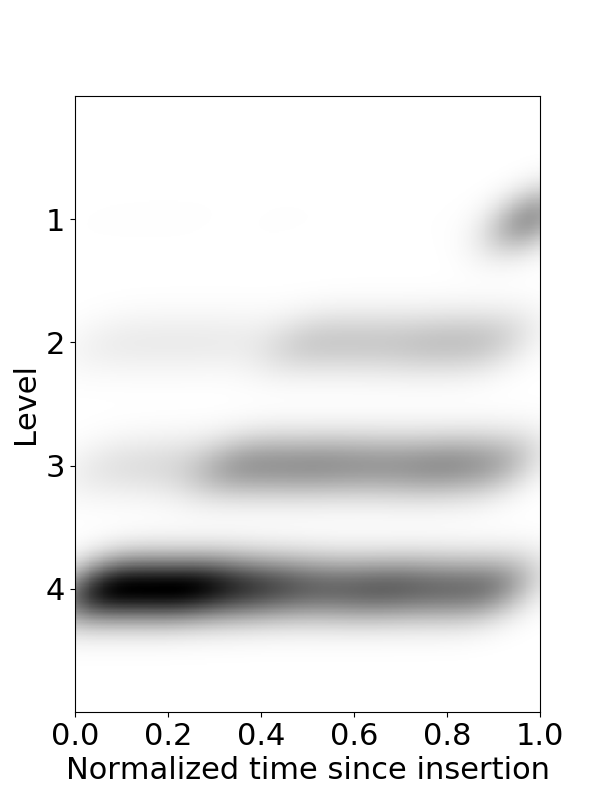}  
  \caption{Compaction prioritized by size (\textit{kByCompensatedSize})}
  \label{fig:time-first}
\end{subfigure}%
~
\begin{subfigure}{0.23\textwidth}
  \centering
  \includegraphics[width=3.6cm,height=3.7cm]{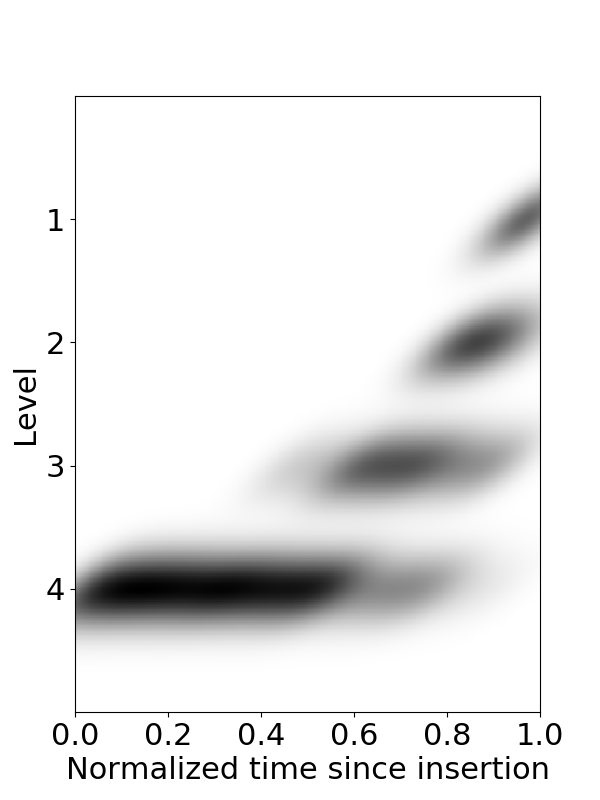}  
  \caption{Compaction prioritized by time (\textit{kOldestSmallestSeqFirst})}
  \label{fig:time-second}
\end{subfigure}
\caption{Distribution of keys across levels based on time}
\label{fig:time}
\end{figure}

The merging process moves data from one level to the next over time.  This puts recent data in the upper levels and older data in the lower levels, providing a natural framework for a lifecycle-aware storage engine proposed in this paper.  In Figure \ref{fig:time}, we present the results of an experiment using RocksDB with an LSM-Tree having five levels (zero through four), with Level-0 starting at 64MB and $T=2$. We inserted data at a steady rate until all the levels were full, with background compaction enabled. We show the distribution of keys in terms of their time-since-insertion for two compaction policies commonly used in RocksDB: \textit{kByCompensatedSize} (Figure \ref{fig:time-first}) prioritizes the largest SST, and \textit{kOldestSmallestSeqFirst} (Figure \ref{fig:time-second}) prioritizes SSTs whose key range has not been compacted for the longest time.  For both compaction priorities, each level has a high density of keys within a given time range.  We will use  time-based compaction priority because it is better at distributing keys based on time since insertion.

A \textit{point query} starts from the most recent data and stops as soon as the search key is found (there may be older versions of this key deeper in the LSM-Tree, but the query only returns the latest version). First, the in-memory skiplists are probed. If the search key has not been found, then the sorted runs on disk are searched starting from Level-0. Within a sorted run, binary search is used to find the SST whose key range includes the key requested by the query.  Then, the index block of this SST is binary-searched to identify the data block that may contain the key. Many LSM-Tree implementations include a bloom filter with each SST, and an SST is searched only if the bloom filter reports that the key may exist.  We assume that the ranges of SSTs, the index blocks of SSTs, and bloom filters fit in main memory and are cached, as illustrated in Figure~\ref{fig:rocksdb}.
For \textit{range queries}, all the skiplists and the sorted runs are scanned to find keys within the desired range.
In many implementations (including RocksDB), range queries are implemented using multiple iterators, which are opened in parallel over each sorted run and the skiplists. Then, similar to a \textit{k-way merge}, keys are emitted in sorted order while discarding old versions.


\subsection{Cost Analysis}

We now explain the cost of LSM-Tree writes, point queries, 
range queries, and space amplification \cite{dostoevsky, monkey, chen}.  We assume that leveling is used for compaction, that sorted runs are not partitioned into SSTs, and that the tree is in a steady state, with all levels full and the volume of inserts equal to the volume of deletes.


Table~\ref{tab:symbols} summarizes the symbols.
Let $N$ be the number of records, $T$ be the size ratio between consecutive levels, and $L$ be the number of levels. Let $B$ denote the number of records in each data page, and let $pg$ denote the number of pages in Level-0. For example, with a 4kB page and 100 bytes per record, $B = 40$; with Level-0 of size 64MB, $pg = 16,000$. Level-0 contains at most $B.pg$ entries, and level $i$ ($i \geq 0$) contains at most $T^i.B.pg$ entries. 
The largest level contains approximately $N.\frac{T-1}{T}$ ($\approx T^L.B.pg$) entries. The total number of levels is given by Equation~\ref{eq:levels}.


\begin{equation}
\footnotesize
\label{eq:levels}
L = \left\lceil \log_T \left(\frac{N}{B.pg}.\frac{T-1}{T}\right) \right\rceil
\end{equation}

\textbf{Write amplification:} Inserted or updated keys are merged multiple times across levels over time, therefore the insert or update I/O cost is measured in terms of write amplification. The worst-case write amplification corresponds to the I/O required to merge an entry all the way to the last level. An entry in level $i$ is copied and merged every time level $i-1$ fills up and is merged with level $i$. This can happen up to $T$ times. Adding this up over $L$ levels, each entry is merged $L.T$ times. Since each page contains $B$ entries, the write cost for each entry across all the levels is $O(\frac{T.L}{B})$.

\textbf{Point queries:} 
The worst-case lookup cost for an existing key is  $O(L)$ without bloom filters because the entry may exist in the last level, requiring access to one block (whose range overlaps with the search key) in each level along the way.  With bloom filters, the average cost of fetching a block from the first $L-1$ levels is $(L-1).fpr$, plus one I/O to fetch the entry from last level, where $fpr$ is the false positive rate of the bloom filter.  In practice, $fpr$ is roughly 1\%, giving an I/O cost of $O(1)$.

\textbf{Range queries:} Let $s$ be the selectivity, which is the number of unique entries across all the sorted runs that fall within the target key range. If keys are uniformly spread across the levels, then in each level $i$, $s/T^{L-i}$ entries will be scanned. With $B$ entries per block, the total number of I/Os is $O(\frac{s}{B}\sum \limits_{i=0}^{L} \frac{1}{T^{L-i}})$. Since the largest level contributes most of the I/O, the cost simplifies to $O(\frac{s}{B})$.

\textbf{Space amplification:} This is defined as $amp = \frac{N}{unq} - 1$, where $unq$ is the number of unique entries (keys). The worst-case space amplification occurs when all the entries in the first $L-1$ levels correspond to updates to the entries in the largest level. The first $L-1$ levels contain $\frac{1}{T}$ of the data. Therefore, $\frac{1}{T}$ of the data in the last level are obsolete, giving a space amplification of $O(\frac{1}{T})$. 

\begin{table}[]
    \centering
    \caption{Summary of terms used in this paper}
    \footnotesize
    \begin{tabular}{|c|l|}
        \hline
        $N$ & number of entries \\ \hline
        $L$ & total number of levels \\ \hline
        $T$ & size ratio between adjacent levels \\ \hline
        $B$ & \# of row style entries in a block \\ \hline
        $B_{ji}$ & \#entries in blocks at CG $j$ at level $i$ \\ \hline
        $pg$ & number of blocks in Level-0 \\ \hline
        $c$ & number of columns \\ \hline
        $s$ & range query selectivity (i.e., \# entries selected)  \\ \hline
        $\Pi$ &  set of projected columns \\ \hline
        $g_i$ & \#column groups at level $i$, $1 \leq g_i \leq c$ \\ \hline
        $\cgsz_{ji}$ & size of $j^{th}$ CG at level $i$ \\ \hline
        $\mathbf{CG_i}$ & CGs at level $i$ \\ \hline
        $E^g_i$ & estimated number of CGs required by a projection \\ \hline
        $E^G_i$ & estimated sum of sizes of CGs required by a projection \\ \hline
    \end{tabular}
    \label{tab:symbols}
\end{table}

\section{Real-Time LSM-Tree Design} \label{sec:hybridLSM}
\subsection{Definitions}
\label{sec:definitions}

\begin{figure*}
\centering
  \includegraphics[width=17cm,height=3cm]{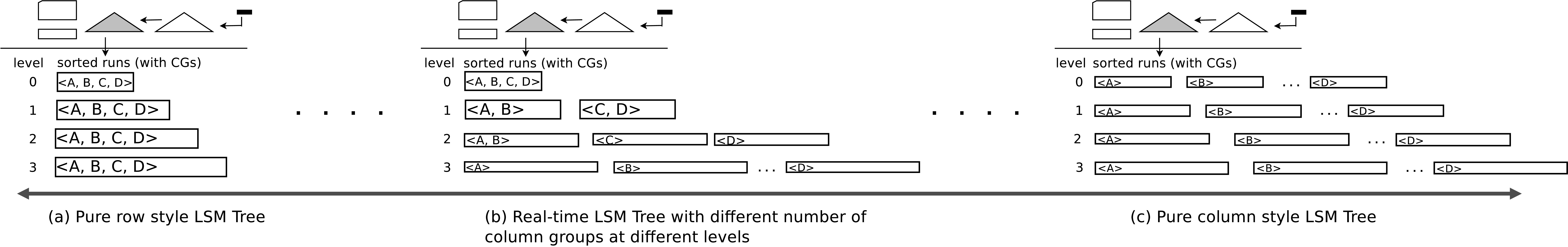}
  \caption{Design space of Real-Time LSM-Trees, with example column group (CG) configurations.}
\label{fig:design-space}
\end{figure*}

\textbf{Lifecycle-driven hybrid workloads:} 
We target HTAP
workloads with high data ingest rates, data volume that requires secondary storage, and access patterns that change with the lifecycle of the data. These workloads include a mix of writes and reads, with recent data accessed by OLTP-style queries (point queries, inserts, updates), and both recent and older data accessed by OLAP-style queries (range queries) \cite{gartnerHTAP}. From a storage engine's viewpoint, we represent these workloads as combinations of inserts, updates, deletes, point reads, and scans. With $key$ as the row identifier, $row$ as the tuple with all the column values, and $\Pi$ as the set of projected columns (e.g., $\Pi = \{A, C\}$ means that the query requires values for columns A and C only), we consider the following operations:
\begin{itemize}[leftmargin=*]
    \item \textit{insert($key$, $row$)}: inserts a new entry.
    \item \textit{read($key$, $\Pi$)}: for the given $key$, reads the values of columns in $\Pi$.
    \item \textit{scan($key_{low}$, $key_{high}$, $\Pi$)}: reads the  values of the columns in $\Pi$ where the key is in the range $key_{low}$, and $key_{high}$. Range queries over non-key columns also use this operator by scanning all the entries and filtering out the entries that are not within the range.
    \item \textit{update($key$, $value_{\Pi}$)}: updates the values of the columns in $\Pi$ for the given $key$. $value_{\Pi}$ contains the column identifiers and their new values. For example, $value_{\Pi} = \{(A,nv_a),$ $ (B,nv_b)\}$ indicates new values for columns A and B for the given key.
    \item \textit{delete($key$)}: deletes the entry identified by $key$.
\end{itemize}

We assume that \textit{read} and \textit{update} access recently inserted keys with a wide $\Pi$ (almost all the columns), while \textit{scan} accesses a range of keys spanning historical and recent data with a narrow $\Pi$ (one column or a few columns depending on the age of the data).


\textbf{Column groups (CGs):} 
A hybrid storage layout is defined by column groups (CGs) that are stored together as rows \cite{blink}. Suppose we have a table with four columns: $A$, $B$, $C$, and $D$.  In a row-oriented layout, there is a single CG corresponding to all the columns.  In a column-oriented layout, each column corresponds to a separate CG.  Other hybrid layouts are possible, e.g., two CGs of <A,B,C> and <D>, where the projection over columns $A$, $B$, and $C$ is stored in row format, and the projection over $D$ is stored separately. 
Column groups are advantageous when some columns are co-accessed often.

\subsection{Design Overview}
\label{sec:hybrid-design}
The insight that makes the Real-Time LSM-Tree a natural fit for a lifecycle-aware storage engine is that 
\textit{different levels may store data in different layouts}.  This creates a design space 
that can be characterized by the column groups used in each level.  In Figure \ref{fig:design-space}, we show three examples.  On the left, we show an extreme
design point corresponding to a row-oriented format, which is used by existing LSM-Tree storage engines.
On the right, we show the other extreme, corresponding to a pure columnar layout.
In the middle, we show a hybrid design, in which Level-0 is row-oriented, levels 1 and 2 use different combinations of CGs, and Level-3 switches to a pure columnar layout. 
This design may be suitable for mixed or HTAP workloads, with the column group configuration depending on the access patterns during the data lifecycle.

In the Real-Time LSM-Tree, we keep the in-memory component and Level-0 the same as in the original LSM-Tree, as described in Section~\ref{sec:background}, to maintain high write throughput. However, the secondary-storage levels beyond Level-0 are split into CGs, where each CG stores its own sorted runs.  As we will see in Section \ref{sec:implementation}, each such sorted run is associated with tail indices and bloom filters to answer queries that access columns within the CG.

Since different levels may have different CG configurations, a Real-Time LSM-Tree must be able to change the data layout as data move from one level to another.  As we will explain in Section \ref{sec:new-compaction}, this can naturally be done during compaction. 

\textbf{CG containment assumption:}
The space of Real-Time LSM-Trees consists of all possible combinations of CGs in each level.  However, we make a simplifying assumption since access patterns throughout the data lifecycle tend to change from row-friendly OLTP to column-friendly OLAP.  We assume that any CG in level $i$ is a subset of (i.e., contained in) a single CG in level $i-1$, for $i\geq1$.  Returning to Figure \ref{fig:design-space}, the design in the middle has two column groups in Level-1: <A,B> and <C,D>.  This means that, for example, a CG of <A,B,C> or a CG of <B,C> is not a valid choice in Level-2. This assumption is not critical to LASER, but it simplifies layout changes during compaction, as we will see in Section \ref{sec:new-compaction}.

\textbf{No replication assumption}: For some workloads, data in a given level may be touched by both OLTP and OLAP style queries, meaning that no single CG layout is suitable for that level. This may be true especially in the last level, which stores the oldest and the majority of the data. For these workloads, a level can be replicated in two layouts, 
at the cost of storage and write amplification. However, we expect such situations to be rare in practice because OLTP patterns tend to be limited to recent data in real-time analytics workloads \cite{htapSurvey, gartnerHTAP}, which are expected to fit in the first few levels. 

\section{LASER Storage Engine} \label{sec:implementation}


We now describe the design of \textit{LASER} -- our HTAP storage engine based on Real-Time LSM-Trees. 
LASER borrows several concepts from column-store systems \cite{abadi-book}: a data model for storing column groups (Section \ref{sec:cg-representation}), column updates (Section \ref{sec:write-ops}), and ``stitching'' individual column values to reconstruct tuples (Section \ref{sec:read-ops}). LASER also requires a mechanism to change the data layout from one level to the next in the Real-Time LSM-Tree (Section \ref{sec:new-compaction}). 

\subsection{Column Group Representation}
\label{sec:cg-representation}
\begin{figure}[t]
\centering
  \includegraphics[width=5cm,height=4cm]{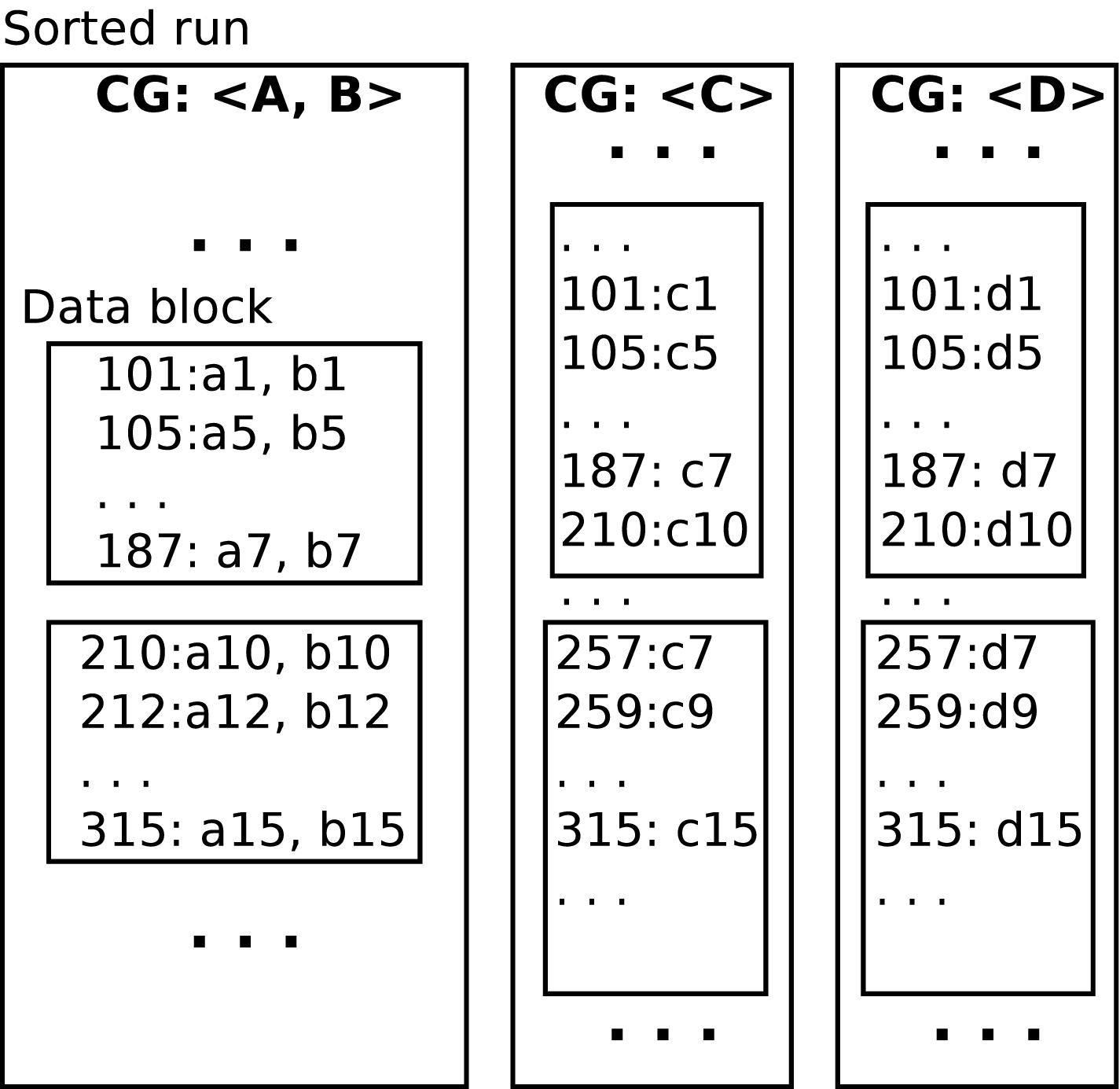}
  \caption{Simulated column-group representation}
\label{fig:cg-layout}
\end{figure}

Since entries in an LSM-Tree are stored across multiple sorted runs and levels, column scans do not access the data contiguously.  To fetch data in sorted order from different levels, we need to locate entries by their keys. Therefore, we store the keys along with the column group values, as shown in Figure \ref{fig:cg-layout}. This is known as simulated columnar storage \cite{abadi-book}, 
and incurs read and storage overhead compared to storing only the column values in a contiguous data block. However, in LSM-Trees, this overhead is reduced due to the leveling merge strategy, and can be further reduced by compressing the data blocks and delta-encoding the keys within each data block. For example, we observed that na\"ively storing keys with column group values took 86GB, using Snappy compression took 51GB, and delta-encoding the keys further reduced the space usage to 48GB. Storing the same amount of data in a pure column-store (MonetDB \cite{monetdb}), which stores only the column values, requires 43GB.

\subsection{Write Operations}
\label{sec:write-ops}

\textbf{Inserts} are \textit{performed in the same way as in original LSM-Trees}, where an entry is inserted in the in-memory skiplist, and is eventually moved to lower levels via flush and compaction jobs. Inserting an existing key (and a corresponding value, containing the values of the remaining attributes) acts as an update, whereas inserting an existing key with a tombstone flag acts as a deletion.

\textbf{Updates} of individual columns may be implemented in two ways.  A straightforward way is to fetch the entire tuple that is to be updated, modify the column being updated, and re-insert the entire tuple.  This is the standard approach in a row-oriented storage engine.
Column-oriented storage engines \cite{vertica,c-store} and some HTAP storage engines \cite{pavlo} allow updates of individual columns.  Similarly, in LASER,  
we allow insertion of partial rows that contain only the updated column values.   
Partial rows are eventually merged with complete rows, or other partial rows, at the time of compaction, and any older column values are discarded. 
For example, suppose we have four columns, <A,B,C,D>, and suppose we  update columns $B$ and $C$ of the tuple with key $100$.  Here, we insert the following key-value pair: $100: -, b', c', -$ where $b', c'$ are the updated values, and $-$ denotes an unchanged value.  If, during compaction, we find another entry for the same key, $100: a, b, c, d$, then the two entries are merged to give $100: a, b', c', d$.  

\begin{figure}[t]
\centering
  \includegraphics[scale=0.2]{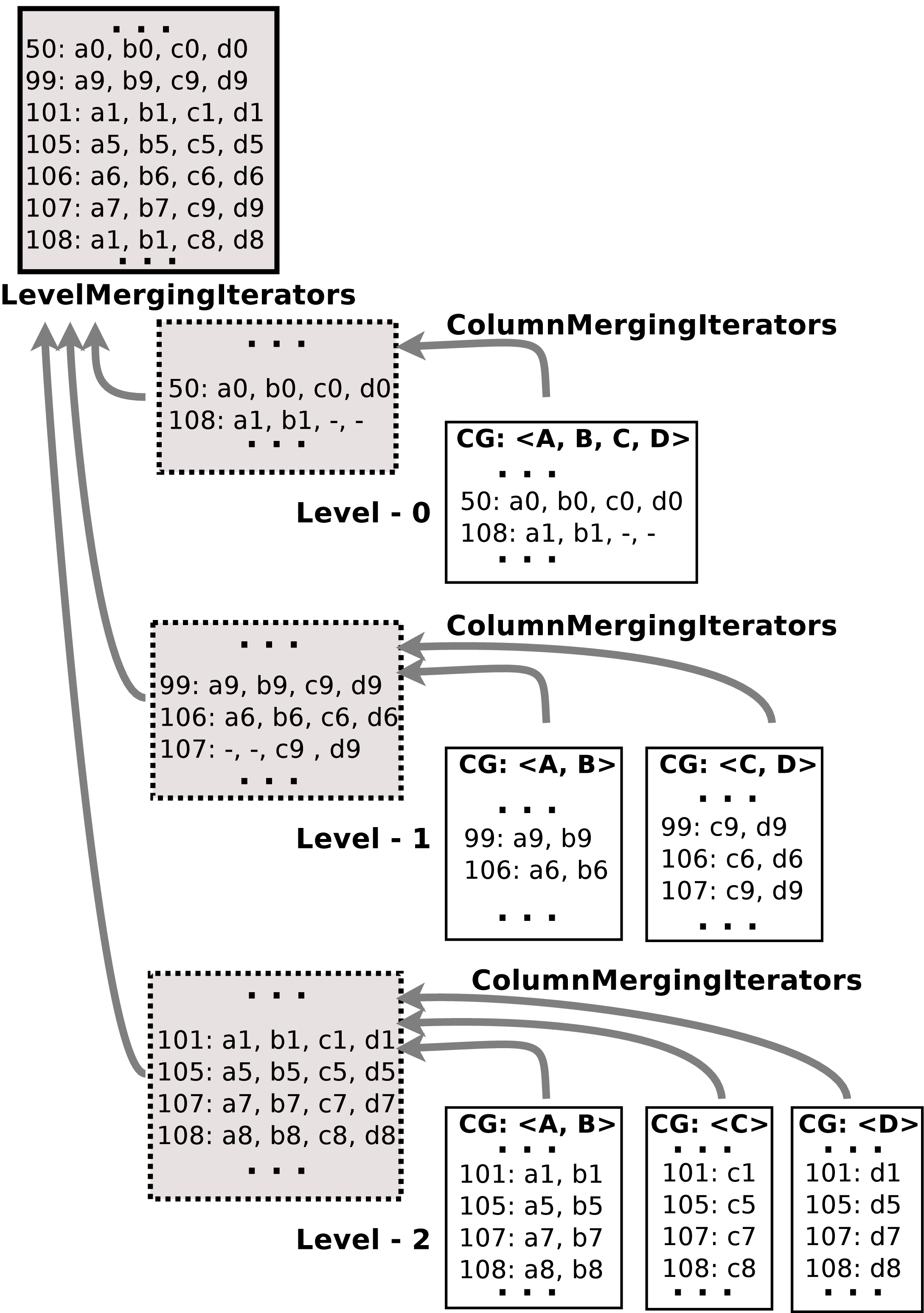}
  \caption{\textit{ColumnMergingIterators} and \textit{LevelMergingIterators}}
\label{fig:iterators}
\end{figure}

\subsection{Read Operations}
\label{sec:read-ops}




\textbf{Point queries (with projections)}  
are handled by searching for the given key in the skiplist, and then down the levels until the latest value is found.  To support projections efficiently, in each level, we only probe the CGs that overlap with the projected columns, and the query result is returned as soon as the values for all of the projected columns are found. Since we allow updates of individual columns, (the latest version of) a given tuple may exist partially in one level and partially in another.  For example, in Figure~\ref{fig:iterators}, the latest values of $A$ and $B$ for tuple 108 exist in Level-0, but the values of $C$ and $D$ exist in Level-2.

\textbf{Range queries (with projections)} 
are also handled by opening iterators for each level and returning values in a sorted order, while discarding older versions of the entries. We optimize range queries with projections by opening iterators only for the overlapping column-groups in each level. As was the case for point queries, subsets of column values may be found across different levels. We use \textit{LevelMergingIterators} to merge values across levels, and to stitch column values within a level we use \textit{ColumnMergingIterators}. We provide the details of these iterators in Section \ref{sec:new-compaction}.

\subsection{Real-Time LSM-Tree Compaction}
\label{sec:new-compaction}
In Section \ref{sec:background}, we described the compaction process used by LSM-Trees to improve query performance.  In LASER, compaction additionally changes the data layout. A compaction job selects a level that overflows the most, and merges its entries with the next level. Using this approach in LASER would require merging entries from all the CGs of an overflowing level with the next level. However, since we allow individual column updates (Section \ref{sec:write-ops}), different CGs can fill up at different rates. For example, some column groups (bank balance, inventory) may be updated more frequently than others (contact information, item description). Therefore, treating all the CGs in the same way when scheduling a compaction job might push certain CG values to deeper levels even when top levels are not full, and therefore disrupt the distribution of entries across levels based on time. We modify the compaction strategy to select the most overflowing CG in the most overflowing level. To determine if a CG is overflowing, we define the capacity of a CG within a level by proportionally dividing the level capacity across all the CGs, and any CG that exceeds its capacity is an overflowing CG.

\begin{figure*}[t]
\centering
  \includegraphics[scale =0.25]{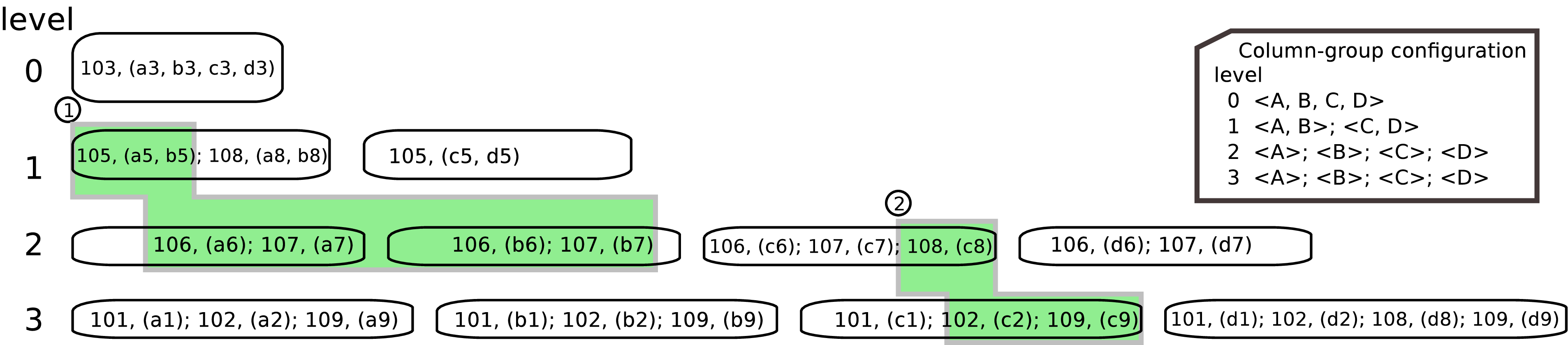}
  \caption{Sorted runs of a Real-Time LSM-Tree with two highlighted compaction jobs. 
  }
\label{fig:hybridLSM}
\end{figure*}

We call this strategy a \textit{CG local compaction} strategy, in which the span of a compaction job is limited to only one CG from level $i$ and the overlapping CGs at level $i+1$. We show two example compaction jobs in Figure \ref{fig:hybridLSM}. Compaction job \circled{1} merges entries from CG <A, B> in level-1 to overlapping CGs (i.e., <A>; <B>) in level-2. Similarly, compaction job \circled{2} is limited to only CG <C> in level-2 and level-3. To perform \textit{CG local compaction}, we require two types of merging iterators: \textit{LevelMergingIterators} that merge entries from different levels, and \textit{ColumnMergingIterators} that combine column values from different CGs within the same level.

\textbf{LevelMergingIterators} 
support range queries and compaction jobs by fetching and merging qualifying tuples from each level, and discarding old attribute values when multiple versions are found for the same key.
Figure~\ref{fig:iterators} shows LevelMergingIterators collecting tuples from three levels to answer a range query for keys between 50 and 108.  Only the latest versions of keys 107 and 108 are returned.

\textbf{ColumnMergingIterators} combine values from different column groups within the same level. For each LevelMergingIterator, multiple ColumnMergingIterators are opened. 
Since there can be only one version for each key and column value within a level, these iterators do not discard old versions. Instead, they fetch all the required column values for each key, some of which may be empty due to column updates. In Figure \ref{fig:iterators}, we show ColumnMergingIterators for each level.  In level-0, the iterators return partial values for 108 because the corresponding entry corresponds to an update of columns $A$ and $B$. Similarly, in level-1, key 107 has a partial value. 

The above iterators are used by \textit{CG local compaction} in the following way: we first identify the most overflowing level, and the most overflowing CG in that level. Then, we identify the overlapping CGs in the next level, open LevelMergingIterators for both levels, and open the required ColumnMergingIterators for the respective LevelMergingIterators. Once the iterators are opened, entries are emitted in sorted order and are written to the new sorted run belonging to the next level.

\section{Cost Analysis} \label{sec:cost-analysis}

In this section, we analyze the cost of each operation supported by LASER, and compare it with the cost of a purely row-oriented LSM-Tree (Section \ref{sec:background}) and a purely column-oriented LSM-Tree (a special case of a Real-Time LSM-Tree with as many CGs as columns). Table \ref{tab:cost-summary} summarizes the operations and their costs. 

We use the variables listed in Table \ref{tab:symbols}.  Let $1 \leq g_i \leq c$ be the number of CGs at level $0 \leq i \leq L$, where $c$ is the number of columns. The size of the $j^{th}$ ($1 \leq j \leq g_i$) CG at level $i$ is defined as the number of columns in the CG and is represented by $\cgsz_{ji}$. $\cgsz_{ji}$ is $c$ for all column-groups at all levels for a row-style LSM-Tree and $1$ for all column-groups at all levels for a column-style LSM-Tree. For each level $i$, we have the following relation between $c, g_i$, and $\cgsz_{ji}$:
\begin{equation}
\small
\label{eq:cgsize}
c = \sum\limits_{j=1}^{g_i} cg\_size_{ji}
\end{equation}

Let $B_{ji}$ be the number of entries in a data block of a $j^{th}$ CG at level $i$. From Section \ref{sec:background}, we know that a row-style LSM-Tree contains $B$ entries in a block. The block size, in bytes, is fixed for an LSM-Tree; for example in RocksDB, it is 4kB by default. If $D$ is the block size in bytes, then we have $D = B.(key$-$size + value$-$size) = B.(1.dt\_size + c.dt\_size)$, where $dt\_size$ is the average datatype size of the columns, which includes the column value and the key. This can be generalized for a Real-Time LSM-Tree, in which a block contains $B_{ji}$ entries: $D = B_{ji}.(1 + cg\_size_{ji}).dt\_size$. For example, in Figure \ref{fig:cg-layout}, the relationship between the number of entries in a block of CG <A,B>, and CG <C> is $D = B_{<A,B>}.(1 + 2).dt\_size = B_{<C>}.(1 + 1).dt\_size$, or $B_{<A,B>} = 2.B_{<C>}/3$. The relationship between $B$ and $B_{ji}$ is as follows.
\begin{equation}
\small
\label{eq:bji}
B_{ji} = B.\frac{(1+c)}{(1+cg\_size_{ji})}
\end{equation}
This gives $B_{ji} = B.(1+c)/2$ for all column-groups at all levels for a column-style LSM-Tree. As $\cgsz_{ji}$ reduces, $B_{ji}$ increases because we can pack more entries of smaller CG size in a block. 

\textbf{Write amplification:} We start with the cost of write amplification for \textit{insert(key, row)} operations. 
For a row-style LSM-Tree, the write amplification was described in Section \ref{sec:background}, i.e.,  $O(T.\frac{L}{B})$. For a column-style LSM-Tree, each level has $c$ column-groups (each with one column).  Therefore, the write amplification is $O(c.T.\frac{L}{B_{ji}})$, where $B_{ji} = B.(1+c)/2$ for all CGs. For a Real-Time LSM-Tree, the write amplification is summed across all the CGs and all the levels. For example, in level-2 of the Real-Time LSM-Tree shown in Figure \ref{fig:hybridLSM}, entries will be merged for CGs <A,B>; <C>; <D> where $B_{02} = B(1+4)/(1+2) = 5B/3$ (i.e., for CG <A,B>) and $B_{12} = B_{22} = 5B/2$. For each CG, the merge cost is given by $T/B_{ji}$ (because entries are merged $T$ times, as explained in Section \ref{sec:background}). The total write amplification cost is: $O(\sum\limits_{i=0}^L \sum\limits_{j=1}^{g_i}T/B_{ji})$. Using Equations \ref{eq:cgsize} and \ref{eq:bji}, this simplifies to $O(\frac{T.L}{B} + \frac{T}{B.c} \sum\limits_{i=0}^{L}g_i)$. The second term (i.e. $\frac{T}{B.c} \sum\limits_{i=0}^{L}g_i$) represents the overhead of storing keys along with CG values due to the simulated column group representation. This overhead is at most $TL/B$ (because $1 \leq g_i \leq c$) in a column-style LSM-Tree.
\begin{equation}
\small
\label{eq:insert}
W := O\left(\frac{T.L}{B} + \frac{T}{B.c}.\sum\limits_{i=0}^{L}g_i \right)
\end{equation}

\textbf{Point lookups:} 
The cost for a row-style LSM-Tree is the same as in Section \ref{sec:background}, i.e. $O(1)$ (assuming the false positive rate of bloom filters is much smaller than 1). For a column-style LSM-Tree, the cost is equal to the number of column groups containing the columns projected by the query. 
For a Real-Time LSM-Tree, this cost is similarly equal to the number column-groups containing the projected columns, summed over all the levels. We use $E^g_i$ ($1 \leq E^g_i \leq g_i$) to define the number of column-groups required at level $i$. 
For example, if there are two CGs, <A, B>; <C, D>, in level $i$, then $E^g_i = 2$ when the projection is $\Pi = \{A, C\}$ and $E^g_i = 1$ when $\Pi = \{A, B\}$. The total I/O cost is therefore:
\begin{equation}
\small
\label{eq:point}
P := O(\sum\limits_{i = 0}^{L}E^g_i)
\end{equation}

\textbf{Range queries:} The I/O cost for a row-style LSM-Tree is the same as in Section \ref{sec:background}, i.e., $O(\frac{s}{B})$.  For a column-style LSM-Tree, this depends on the number of CGs containing the projected columns. Therefore, the I/O cost is $O(|\Pi|.\frac{s}{B.c})$ (here, $B_{ji} = B.(1+c)/2$). For a Real-Time LSM-Tree, different levels contribute different costs depending on the CG configuration. Anytime a CG contains one or more columns projected by the query, the entire block of that CG must be fetched. Therefore, for each level, we have $O(\sum \limits_{j \in G_i} s_i/B_{ji})$, where $s_i$ is the selectivity at level $i$, and $G_i$ is the set of CGs containing the projected columns. In Section \ref{sec:background}, we defined $s$ to be the selectivity of a range query for all the levels; selectivity $s_i$ for an individual level $i$ can be estimated by dividing $s$ by the capacity of that level. Using Equation \ref{eq:bji}, we obtain the following cost for each level: $O(\frac{s_i}{c.B} \sum \limits_{j \in G_i}(1 + \cgsz_{ji}))$. We define $E^G_i := \sum \limits_{j \in G_i} (1 + \cgsz_{ji})$, i.e., the sum of the sizes of all the required CGs and corresponding keys. For example, if there are CGs <A, B>; <C, D> in level $i$, then $E^G_i = 6$ when the projected columns are $\Pi = \{A, C\}$ and $E^G_i = 3$ when $\Pi = \{A, B\}$. The overall cost of a range query is: 
\begin{equation}
\small
\label{eq:range}
Q := O(\sum\limits_{i = 0}^{L} s_i.E^G_i/c.B)
\end{equation}

\textbf{Update amplification:} The update amplification for a row-style LSM-Tree is the same as the insert amplification: $O(\frac{L.T}{B})$. For a column-style LSM-Tree, the cost depends on the number of column values that are updated due to our \textit{CG local compaction} strategy (Section \ref{sec:new-compaction}). The amplification is given by $O(\frac{L.T.|\Pi|}{B.c})$, where $\Pi$ is the set of updated columns. For a Real-Time LSM-Tree, update amplification depends on the sum of the sizes of the required CGs. This is estimated by $E^G_i$ (see range query cost above). Therefore, the amplification of an update operation is
\begin{equation}
\small
\label{eq:update}
U := O(\sum\limits_{i = 0}^{L} T.E^G_i/c.B)
\end{equation}

\textbf{Space amplification:} As explained in Section \ref{sec:background} for a row-oriented LSM-Tree, the worst-case space amplification in a Real-Time LSM-Tree happens when the first $L-1$ levels contain updates of entries in the last level. The fraction of entries in the first $L-1$ levels is still $\frac{1}{T}$. Therefore, the space amplification is still $O(\frac{1}{T})$.

\begin{table}[]
\small
    \centering
\begin{tabular}{ p{1.8cm} | p{1.2cm} | p{2.5cm} | p{1.4cm} } 
Operation & Row-style LSM-Tree & Real-Time LSM-Tree & Column-style LSM-Tree \\ \hline
\\[-0.8em]
Insert amplification (W) & $O(\frac{T.L}{B})$ & $O(\frac{T.L}{B} + \frac{T.\sum\limits_{i=0}^{L}g_i}{B.c} )$ & $O(\frac{T.L}{B})$ \\
Existing key lookup (P) & $O(1)$ & $O(\sum\limits_{i = 0}^{L}E^g_i)$ & $O(|\Pi|)$ \\
Range query (Q) & $O(\frac{s}{B})$ & $O(\sum\limits_{i = 0}^{L}\frac{s_i.E^G_i}{c.B})$ & $O(\frac{|\Pi|.s}{c.B})$ \\ 
Update amplification (U) & $O(\frac{T.L}{B})$ & $O(\sum\limits_{i = 0}^{L}\frac{T.E^G_i}{c.B})$ & $O(\frac{T.L.|\Pi|}{c.B})$ \\ 
\end{tabular}
\caption{Summary of operations and their costs.}
\label{tab:cost-summary}
\end{table}

\section{Design Selection} \label{sec:best-design}

We now describe how to select a suitable Real-Time LSM-Tree design for a given workload using the cost analysis from Section \ref{sec:cost-analysis}.
Our goal is to find an optimal CG configuration for each level to minimize the total I/O cost for a given workload.

In the context of LSM-trees, this problem is critically different due to the flexibility of assigning a different layout for each level of the tree. That is, we are not searching for a single CG layout across the whole data and tree, but rather we are searching for an optimal layout for each level of the tree in a way that holistically optimizes the overall performance. A critical invariant is the CG containment constraint described in Section \ref{sec:hybrid-design}, (a CG at level $i$ must be a subset of some CG at level $i-1$). LASER makes a decision per level regardless of whether the LSM-tree is based on leveling (one run per level), tiering, or lazy leveling (where there may be several runs per level). This is because in the latter case, runs overlap in terms of the range of values stored, and so all runs will see the same access patterns given a workload at this level.

In the remainder of this section, we define the optimization problem in the context of Real-Time LSM-trees, and we describe our search algorithm, which is inspired by Hyrise \cite{hyrise} and brings novel design elements to allow different layouts in every level of the tree and to ensure CG containment.



\subsection{Input}

\textbf{Parameters:} To find an optimal CG layout for a given workload, LASER requires: 1) parameters defining the Real-Time LSM-Tree structure, and 2) parameters defining the workload. As explained in Section \ref{sec:cost-analysis}, the costs of the operations depend on the Real-Time LSM-Tree structure, which is defined by the parameters $T$, $L$, and $B$ (Section \ref{sec:background}), and on the CG configuration $\mathbf{CG}$. We represent the workload by $\mathbf{wl}$, which is a set of operations. Let $w$ be the number of \emph{insert} operations, $p$ be the number of \emph{read} operations for existing keys, $q$ be the number of \emph{scan} operations, and $u$ be the number of \emph{update} operations in $\mathbf{wl}$. Since we are searching for an optimal CG layout for each level independently, we additionally define level $i$'s workload by $\mathbf{wl_i}$, and similarly, $p_i$, $q_i$, and $u_i$ represent the number of \emph{read}, \emph{scan}, and \emph{update} operations, respectively, served at level $i$.

\textbf{Obtaining parameter values:} We assume that the values of the LSM-Tree parameters ($T$, $L$, $B$) are fixed based on the data size ($N$) and the operating system configuration (e.g., page size). 
Past research showed how to tune T and L 
in an LSM-tree \cite{wacky, monkey, dostoevsky}. Furthermore, B is usually fixed based on a 4kB block size (as in RocksDB). Overall, these parameter choices are orthogonal to LASER: they govern the high-level LSM-tree architecture while LASER optimizes the architecture within each run.
As for the workload, we assume that, at the logical level, it consists of SQL statements. For the LASER storage engine, we convert the workload to the operations 
defined in Section \ref{sec:definitions}. Profiling the workload $\mathbf{wl_i}$ at each level allows us to determine the values for $w$, $p_i$, $q_i$, $u_i$, and $s_i$. Finally, the values for $E^g_i$ and $E^G_i$ are determined by the workload trace and the CG configuration under consideration, as discussed in Section \ref{sec:cost-analysis}.

\subsection{Optimization Problem}
\textbf{Cost function:} 
Let $W_k$ be the cost of the $k^{th}$ write operation in the workload, obtained using Equation~\ref{eq:insert}; we define $P_k$, $Q_k$ and $U_k$ similarly based on Equations~\ref{eq:point} through \ref{eq:update}.  Following previous work on LSM-Tree design \cite{dostoevsky}, we compute the cost of a workload for a given CG configuration $\mathbf{CG}$ by adding up the costs of each operation, as shown in Equation \ref{eq:cost}.
\begin{equation}
\small
\label{eq:cost}
cost(\mathbf{CG}) := \sum \limits_{k=1}^wW_k + \sum \limits_{k=1}^pP_k + \sum \limits_{k=1}^qQ_k + \sum \limits_{k=1}^uU_k
\end{equation}
Since we need to find an optimal CG configuration at each level using per-level workload statistics, the cost function in Equation \ref{eq:cost} can be split into per-level cost, given by the following equation:
\begin{equation}
\small
\label{eq:level-cost}
cost(\mathbf{CG_i}) := \frac{w.T.g_i}{B.c} + \sum \limits_{k=1}^{p_i}E^g_{ik} + \sum \limits_{k=1}^{q_i} \frac{s_{ik}.E^G_{ik}}{c.B} + \sum \limits_{k=1}^{u_i} \frac{T.E^G_{ik}}{c.B}
\end{equation}
Here, $\mathbf{CG_i} = \{cg_{i1}, cg_{i2}, ..., cg_{ig} \}$ is the partitioning of columns into $g$ groups at level $i$ that satisfies the CG containment constraint.

\textbf{Optimization problem:} 
For each level $i$, we want to find an optimal $\mathbf{CG_i}$ such that $cost(\mathbf{CG_i})$ is minimized for the workload $\mathbf{wl_i}$ and the CG containment constraint is satisfied. This leads to the following optimization problem:
\begin{align}
\small
    \label{eq:opt}
    &\forall i: 1 \leq i \leq L \\
    &\mathbf{CG^*_i} = \arg\!\min_{\mathbf{CG_i}}\, cost(\mathbf{CG_i}) \nonumber \\
    &s.t. \forall cg_{ij} \in \mathbf{CG_i} \ \exists \ cg_{(i-1)k} \in \mathbf{CG_{(i-1)}} \ | \ cg_{ij} \subseteq cg_{(i-1)k}   \nonumber
\end{align}
Recall that we keep level-0 row-oriented, so the CG containment constraint is trivially satisfied for level-1.

\subsection{Our Solution}
\label{sec:best-design-sol}
Previous work \cite{hyrise} takes the following three-step approach: 1) pruning the space of candidate CGs, 2) merging candidate CGs to avoid overfitting, and 3) selecting an optimal CG layout from the candidate CGs. The \textit{CG containment} constraint can be added to the first step, further pruning the space of candidate CGs. 

Let $\{a_1, a_2, ..., a_c\}$ be the attributes in relation $\mathbf{R}$, and let $\Pi_{j}$ be the projection of the $j^{th}$ operation (point lookup, range query, or update operation) at level $i$. In the first step, we generate a CG partitioning with the smallest subsets, where every subset contains columns that are co-accessed by at least one operation. This is done by recursively splitting the attributes of $\mathbf{R}$ using the projections $\Pi_{j}$. For example, suppose $\mathbf{R} = \{a_1, a_2, a_3, a_4\}$, and let $\Pi_1 = \{a_2, a_3, a_4\}$, $\Pi_2 = \{a_1, a_2 \}$, and $\Pi_3 = \{a_1, a_2, a_3, a_4\}$. Then, splitting using $\Pi_1$ gives subsets: $\{a_1\}, \{a_2, a_3, a_4\}$, and further splitting using $\Pi_2$ gives subsets: $\{a_1\}, \{a_2\}, \{a_3, a_4\}$ ($\Pi_3$ does not split any subsets). 

The next step is to merge the subsets from the previous step. This is beneficial for point queries, which typically have wider projections, while smaller subsets are beneficial for range scan operations, which typically have narrow projections. This tension between the access patterns of point queries and scan operations is used to decide which subsets should be merged. We merge smaller subsets only if the cost of running the workload with the larger subsets is lower. To systematically evaluate all merging possibilities, we start with the smallest subsets from the previous step, and consider all possible permutations of them for merging. 

Finally, in the third step, we generate all possible CG partitions (covering all attributes of $\mathbf{R}$) from the subsets generated in the previous step, and output the least-cost solution (Equation \ref{eq:level-cost}).

To satisfy the CG containment constraint, when considering level $i$, we change the initial set of attributes $\mathbf{R}$ to be the set of attributes from one CG at level $i-1$, and we separately execute our solution for each CG at level $i-1$. For example, if level-2 has CGs: $<a_1, ..., a_4>$; $<a_5, ..., a_8>$, then we solve two CG selection problems for level-3, one with $\mathbf{R} = \{a_1, ..., a_4\}$ and one with $\mathbf{R} = \{a_5, ..., a_8\}$. 
The design selection algorithm starts with level-1, where the complete schema $\mathbf{R}$ is split into CGs using the three steps described above. Then, this process is repeated for level-2 onwards, where each CG at level $i-1$ is split into optimal CGs for level $i$.
The worst case time complexity of finding an optimal CG configuration at a single level is given by \cite{hyrise}, which is exponential in the number of partitions generated in the first step. The overall worst case time complexity for all levels equals the number of levels times the worst case complexity of each level. Since the number of partitions is small in practice \cite{hyrise} and the CGs get smaller from one level to the next, the actual time taken by the design selection algorithm is expected to be small.  For example, in our evaluation (Section \ref{sec:evaluation}), design selection took only 3 seconds for 100 columns and 8 LSM-Tree levels.



\section{Evaluation}
\label{sec:evaluation}
We now show that: 
    1) the empirical behaviour of LASER matches the cost model from Section \ref{sec:cost-analysis},
    2) LASER outperforms pure row-store, pure column-store, and other column-group hybrid designs, and
    3) LASER is robust to minor workload changes.



\textbf{Setup:} We deployed LASER on a Linux machine running 64-bit Ubuntu 14.04.3 LTS. The machine has 12 CPUs across two NUMA nodes (Intel Xeon E5-2603 v3 @ 1.60GHz), with 15MB of L3 cache, 16GB of RAM, and a 4TB hard drive (Seagate ST4000NM0033, Serial, SATA Rev 3.0).

\textbf{Implementation:}
We implemented LASER on top of RocksDB 5.14. We added the components described in Section \ref{sec:implementation}: simulated CG layout, CG updates, support for projections in queries, \textit{LevelMergingIterators} and \textit{ColumnMergingIterators}, and the CG local compaction strategy.  We reused other necessary but orthogonal components provided by RocksDB, such as in-memory skiplists, index blocks for SSTs, bloom filters, snapshots, and concurrency. To collect workload traces for design selection, we modified the RocksDB profiling tools to collect per-level statistics about operations and their projections. We implemented the design selection algorithm as an offline process that takes in the workload trace and the LSM-Tree parameters as input. 

\textbf{Configuration:} Unless specified otherwise, we use leveling compaction with \textit{kOldestLargestSeqFirst} compaction priority, with up to 6 compaction threads. 
We use the RocksDB default values of other parameters such as Level-0 size, SST size, and compression.

\textbf{Compaction:}
While we use leveling, the results are orthogonal to the compaction strategy: the number of entries in every level remains constant given a fixed size ratio (T). For example, tiering, the write optimized merging strategy, or lazy leveling and the wacky continuum \cite{wacky}, which balance read and write costs, only affect the number of runs within a level, not the number of entries in a level (since runs will simply be smaller with those strategies compared to leveling). In our experiments, we vary the size ratio (T), which affects how entries spread across the levels and the number of levels. This is critical as it affects the number of column-group layouts a Real-Time LSM-tree can hold simultaneously.

\textbf{Workload:}
We generate workloads using the benchmark proposed by previous work on HTAP systems \cite{pavlo,stratos2019}. The benchmark consists of transactional and analytical queries common in HTAP workloads: ($Q_1$) \textit{inserts} new tuples, ($Q_2$) is a \textit{point query} that selects a specific row, ($Q_3$) is an \textit{update query} that updates a subset of attributes of a specific row, ($Q_4$) is an \textit{arithmetic query} that sums a subset of attributes over the selected tuples, and ($Q_5$) is an \textit{aggregate query} that computes the maximum values of selected attributes over selected tuples. These queries are written in SQL as follows: 

Transactional: \\
$Q_1$: \textbf{INSERT INTO} R \textbf{VALUES} ($a_0, a_1, ..., a_c$) \\
$Q_2$: \textbf{SELECT} $a_1, a_2, ..., a_k$ \textbf{FROM} R \textbf{WHERE} $a_0 = v$\\
$Q_3$: \textbf{UPDATE} R \textbf{SET} $a_1 = v_1, ..., a_k = v_k$ \textbf{WHERE} $a_0 = v$\\

Analytical: \\
$Q_4$: \textbf{SELECT} $a_1 + a_2 + ... +a_k$ \textbf{FROM} R \textbf{WHERE} $a_0 \in [v_s, v_e)$\\
$Q_5$: \textbf{SELECT} $MAX(a_1), ..., MAX(a_k)$ \textbf{FROM} R \textbf{WHERE} $a_0 \in [v_s, v_e)$\\

The parameters $k$, $v$, $v_s$, and $v_e$, control projectivity, selectivity, overlap between queries, and access patterns throughout the data lifecycle.
The benchmark includes two types of tables: narrow (30 columns) and wide (100 columns). Each table contains tuples with a 8-byte integer primary key $a_0$ and a payload of $c$ 4-byte integer columns ($a_1, a_2, ..., a_c$). Unless otherwise noted, we use the table with 30 columns, with uniformly distributed integer values as keys. In all  experiments, we run an initial data load phase, followed by a steady workload phase in which we record measurements. 

Note that we do not use complex OLAP queries from benchmarks such as TPC-H and TPC-DS since our focus is on the performance of storage engine operations (inserts, updates, point lookups and scans) rather than query optimizations.  Furthermore, we do not test queries with predicates on non-key columns since LASER converts these to a sequential scan, as described in Section~\ref{sec:definitions}.  Secondary indices can speed up these types of queries, but this is orthogonal to the data layout issues we study in LASER.

\begin{figure*}[t]
\begin{subfigure}[b]{0.2\textwidth}
  \includegraphics[width=4cm,height=2.7cm]{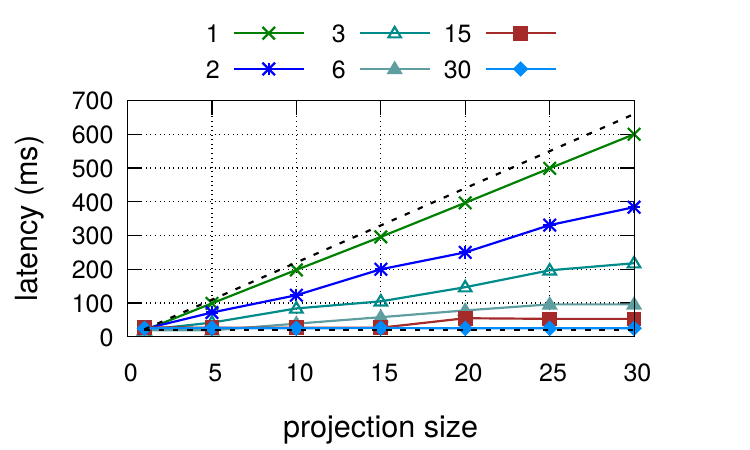}  
\end{subfigure}%
~
\begin{subfigure}[b]{0.2\textwidth}
  \includegraphics[width=4cm,height=2.7cm]{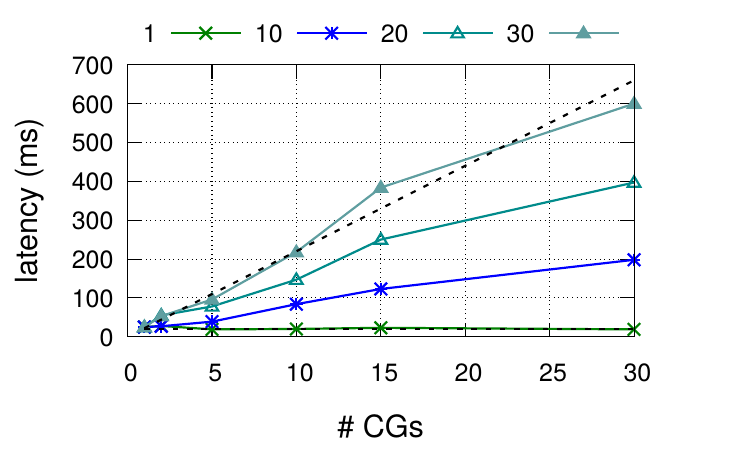}  
\end{subfigure}%
~
\begin{subfigure}[b]{0.2\textwidth}
  \includegraphics[width=4cm,height=2.7cm]{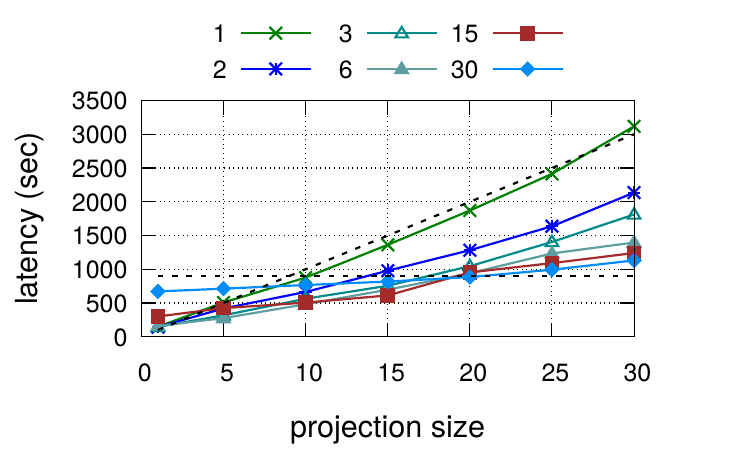}  
\end{subfigure}%
~
\begin{subfigure}[b]{0.2\textwidth}
  \includegraphics[width=4cm,height=2.7cm]{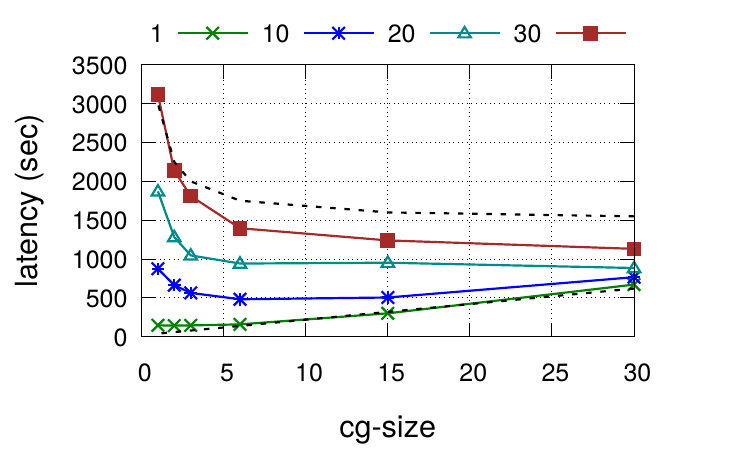}  
\end{subfigure}%
~
\begin{subfigure}[b]{0.2\textwidth}
  \includegraphics[width=4cm,height=2.6cm]{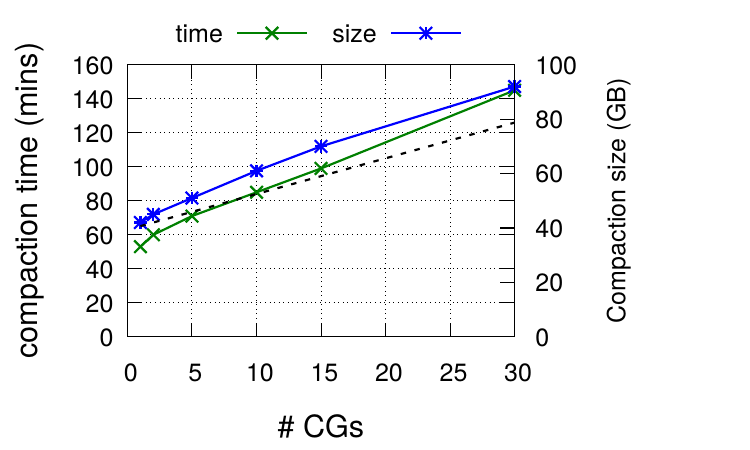}
\end{subfigure}

\begin{subfigure}[t]{0.2\textwidth}
  \includegraphics[width=4cm,height=2.7cm]{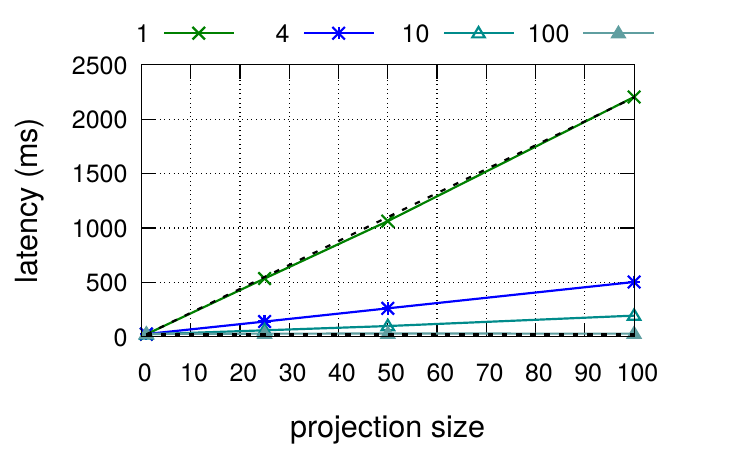}  
  \caption{\small{\textit{Read}: average latency w.r.t. projection size for \\ different CG sizes}}
  \label{fig:exp1-read-pj}
\end{subfigure}%
~
\begin{subfigure}[t]{0.2\textwidth}
  \includegraphics[width=4cm,height=2.7cm]{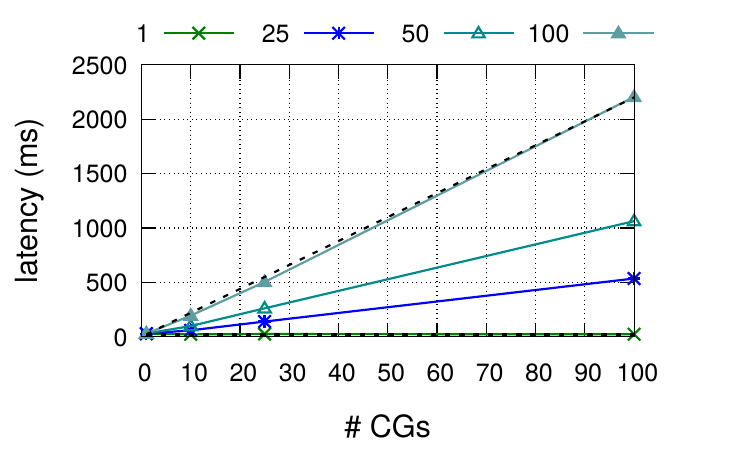}  
  \caption{\small{\textit{Read}: average latency w.r.t. \#CGs for different \\projection sizes}}
  \label{fig:exp1-read-cg}
\end{subfigure}%
~
\begin{subfigure}[t]{0.2\textwidth}
  \includegraphics[width=4cm,height=2.7cm]{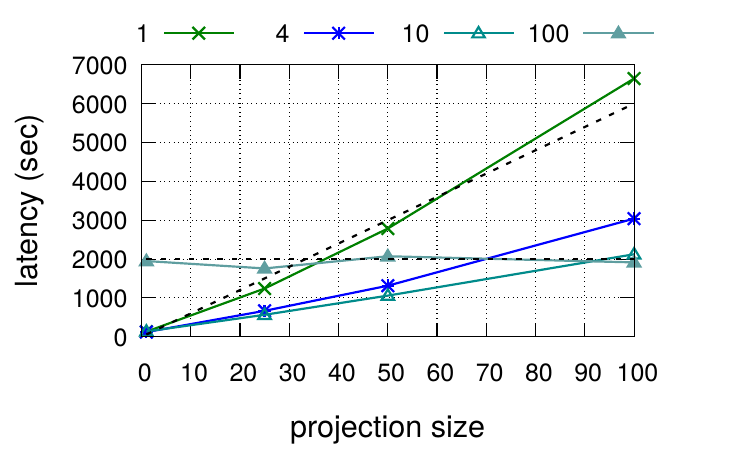}  
  \caption{\small{\textit{Scan}: average latency w.r.t. projection size \\ for different CG sizes}}
  \label{fig:exp1-scan-pj}
\end{subfigure}%
~
\begin{subfigure}[t]{0.2\textwidth}
  \includegraphics[width=4cm,height=2.7cm]{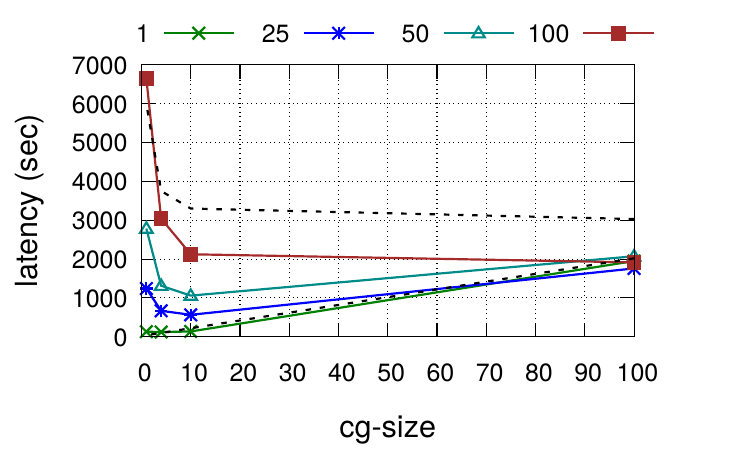}  
  \caption{\small{\textit{Scan}: average latency w.r.t. CG sizes for \\ different projection sizes}}
  \label{fig:exp1-scan-cg}
\end{subfigure}%
~
\begin{subfigure}[t]{0.2\textwidth}
  \includegraphics[width=4cm,height=2.6cm]{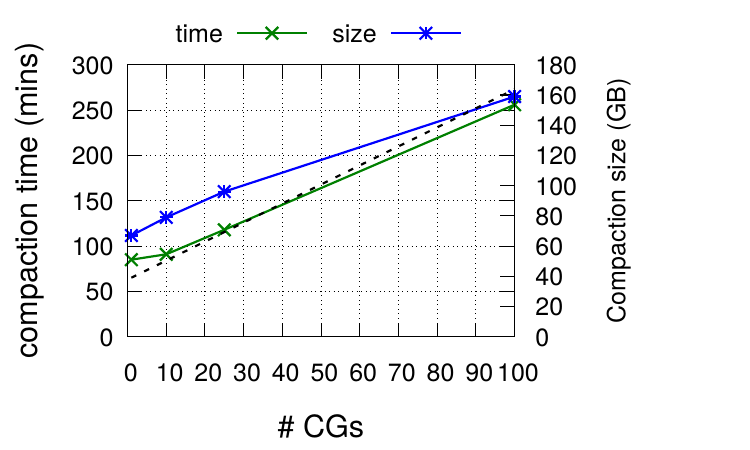}
  \caption{\small{\textit{Write amplification:} compaction time and size w.r.t. \# CGs}}
  \label{fig:exp1-compaction-cg}
\end{subfigure}
\caption{The cost of operations in LASER matches the cost model  in Section \ref{sec:cost-analysis}. The top row corresponds to the narrow table with T=2, and the bottom row corresponds to the wide table with T=10.}
\label{fig:exp1}
\end{figure*}

\vspace{-2pt}
\subsection{Validation of Cost Model}
\label{sec:cg-microbench}
\textbf{Goal:} We begin by validating the costs of point reads, range scans, and write amplification presented in Section \ref{sec:cost-analysis}. 

\textbf{Methodology:}
For a fixed schema and parameters (i.e., $c$ and $B$), the cost of these operations depends on the query projection size and the CG configuration. We validate the cost model using the narrow table and $T=2$, 
as well as the wide table with $T=10$.
For the narrow table, we consider six Real-Time LSM-Tree designs, in which the CG sizes vary from 1 to 30, covering the pure row and pure column layouts, and other designs in between. For each design, we use $g = 30/cg\_size$ equi-width column groups in each level, and we set $cg\_size$ to a value in $\{1, 2, 3, 6, 15, 30\}$. In each design, the LSM-Tree has 8 levels with Level-0 in row-format. 
For the wide table, we consider four Real-Time LSM-Tree designs, with $cg\_size$ values in $\{1, 4, 10, 100\}$, and five LSM-Tree levels.
To generate \textit{read} and \textit{scan} operations, we use $Q_2$ and $Q_5$, respectively, and we vary $k$ from one to 30 to control the projection size. 
Queries are executed after the load phase (400 million entries loaded into the narrow table, and 200 million entries into the wide table) with the OS cache cleared, and we measure the average latency. The write amplification cost is reflected in the background compaction process. To measure compaction time, we load all the entries in Level-0, with compaction disabled, and then schedule compaction manually and measure its runtime. Compaction ends when no level exceeds its capacity. Compaction size is measured as the total bytes written.

\textbf{Results:}
Figures \ref{fig:exp1-read-pj} and \ref{fig:exp1-read-cg} show the latency of \textit{read} operations w.r.t.\ the projection size and the number of CGs, respectively. The top figures correspond to the narrow table and the bottom figures correspond to the wide table. In Figure \ref{fig:exp1-read-pj}, when the CGs are small (similar to a column-oriented layout), latency increases linearly with the projection size because more CGs are fetched from disk. When the CGs are large (similar to a row-oriented layout), latency stays unchanged with the projection size because for any projection size, all the columns are fetched. This is also implied by the point query cost in Equation \ref{eq:point}, which is plotted in black dotted lines for cg\_size=1 (top line) and for cg\_size=30/100 (bottom line). The empirical data in Figure \ref{fig:exp1-read-pj} thus agree with the cost equation.

In Figure \ref{fig:exp1-read-cg},  we vary the number of CGs while keeping the projection sizes fixed.  For wide projections (i.e., fetching complete rows), the cost increases linearly with the number of CGs, because each CG is fetched in a separate disk I/O. However, for narrow projections (i.e., fetching a single column value), the I/O cost stays unchanged because a single disk I/O is enough to fetch the required column value. This is consistent with the point query cost given by Equation \ref{eq:point}, which is plotted in black dotted lines for projection size 30/100 (top line) and 1 (bottom line) in Figure \ref{fig:exp1-read-cg}.

In Figures \ref{fig:exp1-scan-pj} and \ref{fig:exp1-scan-cg}, we measure the latency of \textit{scan} operations w.r.t.\ the projection size and CG size, respectively. Again, the top figures correspond to the narrow table and the bottom figures correspond to the wide table. Similar to Figure \ref{fig:exp1-read-pj}, we vary projection size in Figure \ref{fig:exp1-scan-pj}. For small CGs (similar to a column-oriented layout), latency increases linearly with the projection size because more disk I/O is required to fetch more CGs. However, for large CGs (similar to a row-oriented layout), latency stays almost constant with projection size, because many columns are fetched in a single disk I/O. This is consistent with the range query cost given by Equation \ref{eq:range}, which is plotted in black dotted lines for CG size 1 (top line) and 30/100 (bottom line) in Figure \ref{fig:exp1-scan-pj}. 

In Figure \ref{fig:exp1-scan-cg}, we vary the CG size while keeping the projection sizes fixed. For wider projections, latency should stay constant with CG size, because almost all the columns are fetched irrespective of the CG layout.
However, latency decreases as CG size increases because of the simulated CG layout used in LASER. For large CGs, we fetch the key only once, whereas for smaller CGs, the key is fetched along with each CG, which increases latency. The change in latency for wider projections is proportional to $const_1/cg\_size + const_2$ (Equation \ref{eq:range}), as shown by the top black dotted line in the Figure \ref{fig:exp1-scan-cg}. For smaller projections, we expect latency to increase with CG size because of the overhead of fetching unnecessary columns. This is reflected in Figure \ref{fig:exp1-scan-cg} and matches the cost in Equation \ref{eq:range}.
Similar observations were made in prior work on HTAP systems that allow configurable column groups \cite{h20, pavlo}. 

In Figure \ref{fig:exp1-compaction-cg}, we show that the time and size of compaction jobs for different CG sizes matches our write amplification cost (Equation \ref{eq:insert}), shown using a black dotted line for reference. 
\begin{figure*}[ht]
\begin{subfigure}[t]{\textwidth}
  \centering
  \includegraphics[width=12cm,height=4.4cm]{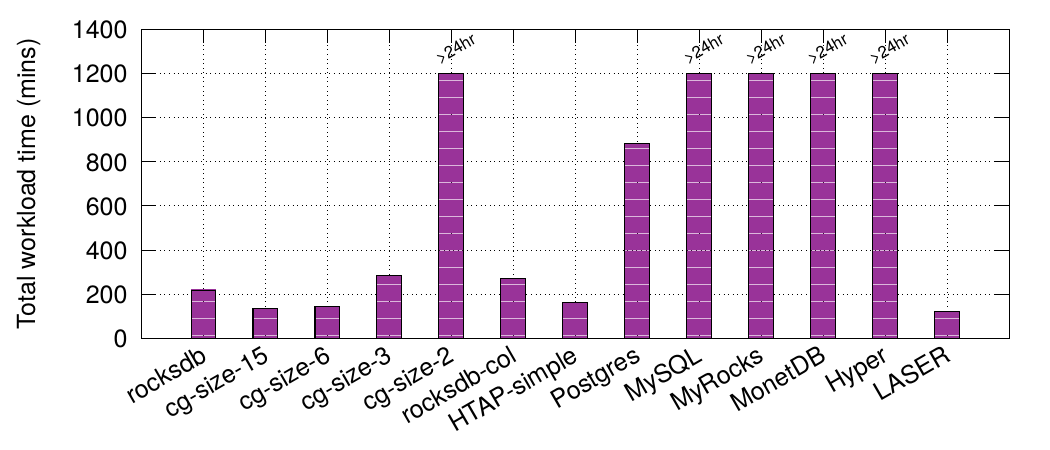}  
  \caption{Workload runtime of different designs}
  \label{fig:exp4-total-time}
\end{subfigure}
\begin{subfigure}[t]{\textwidth}
\centering
  \includegraphics[width=12cm,height=4.4cm]{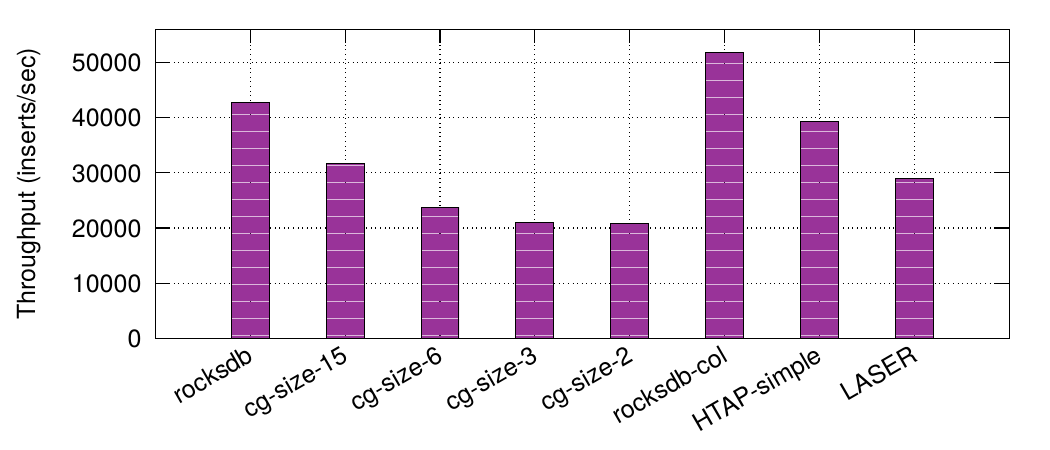}
  \caption{Insert throughput during load phase}
  \label{fig:exp4-insert-throughput}
\end{subfigure}
\begin{subfigure}[t]{\textwidth}
  \centering
  \includegraphics[width=14cm,height=4.4cm]{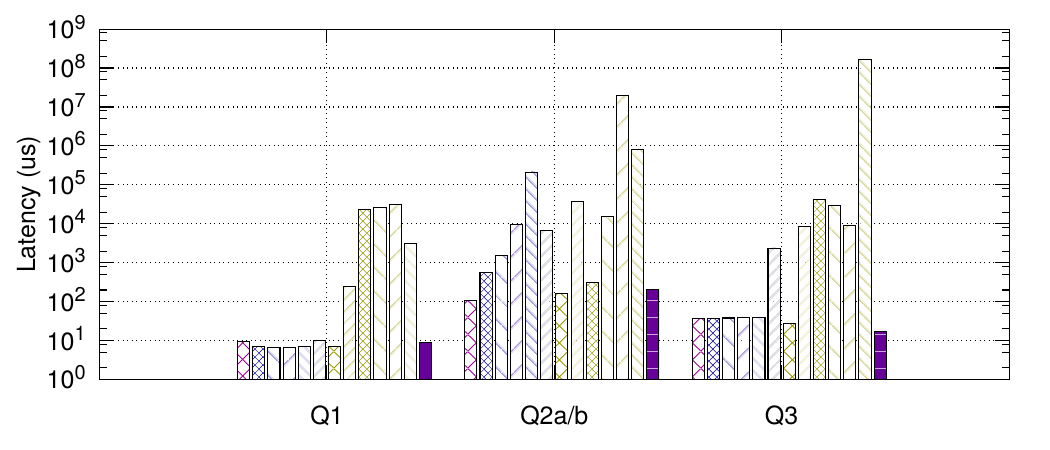}  
  \caption{Latency of inserts (Q1), point queries (Q2a, Q2b), and updates (Q3).}
  \label{fig:exp4-read}
\end{subfigure}
\begin{subfigure}[t]{\textwidth}
  \centering
  \includegraphics[width=14cm,height=4.4cm]{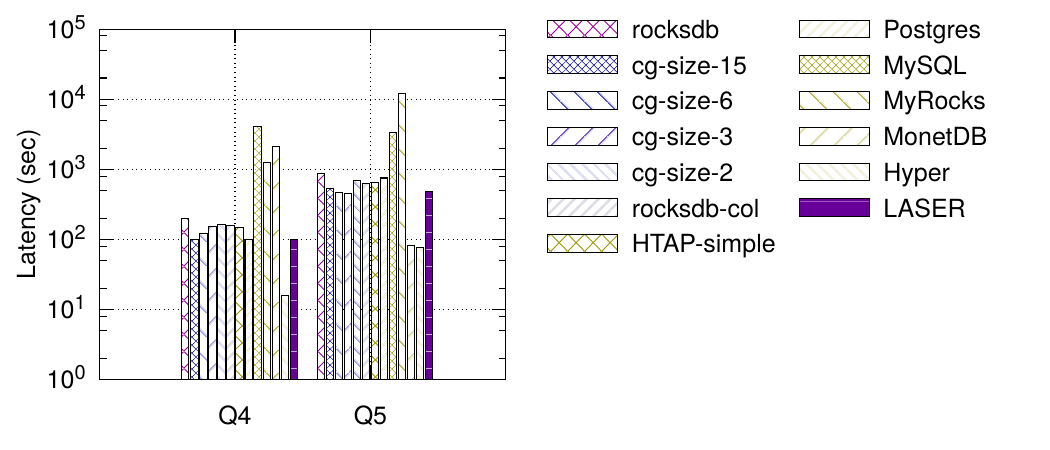}  
  \caption{Latency of range queries (Q4, Q5).}
  \label{fig:exp4-scan}
 \end{subfigure}%
\caption{LASER performs the best on the HTAP workload (HW).}
\label{fig:exp4}
\end{figure*}

\begin{figure}
\begin{subfigure}{0.5\textwidth}
  \centering
  \includegraphics[scale=0.25]{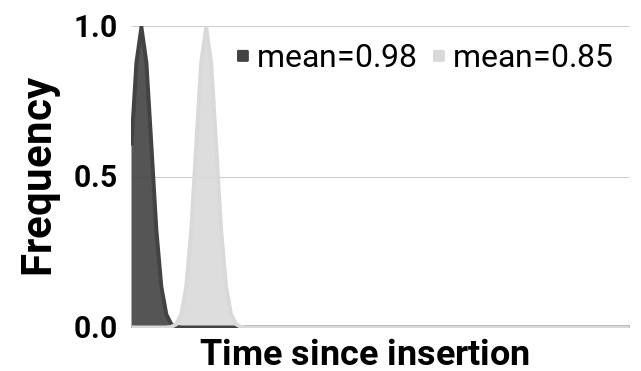}  
  \caption{{$Q_{2a}$: mean=0.98, \\ $Q_{2b}$: mean=0.85}}
  \label{fig:exp4-norm98}
\end{subfigure}
\begin{subfigure}{0.5\textwidth}
\centering
  \begin{tabular}{|l|}
        \hline
        \scriptsize \makecell{$L0: <1$-$30>$ \\
                                $L1: <1$-$30>$ \\
                                $L2: <1$-$15><16$-$30>$ \\
                                $L3: <1$-$15><16$-$30>$ \\
                                $L4: <1$-$15><16$-$20><21$-$30>$ \\
                                $L5: <1$-$15><16$-$20><21$-$30>$ \\
                                $L6: <1$-$15><16$-$20><21$-$27><28$-$30>$ \\
                                $L7: <1$-$15><16$-$20><21$-$27><28$-$30>$} \\ \hline
        
    \end{tabular}
    \caption{Design \textit{D-opt} used by LASER}
    \label{tab:exp-designs}
\end{subfigure}

\caption{\textit{Read} patterns and optimal design used in Sec.\ \ref{sec:exp4}}
\label{fig:exp2-norm}
\end{figure}

\subsection{Performance of LASER}
\label{sec:exp4}
\textbf{Goal:}
We show that LASER can speed up mixed workloads that change with the data lifecycle. We compare 
LASER with a pure row-oriented layout, a pure column-oriented layout, a simple HTAP layout, and various fixed column-group layouts. We also compare LASER with some popular DBMSs from the following categories: row-store DBMSs such as Postgres, MySQL and MyRocks \cite{myrocks}, a column-store DBMS: MonetDB \cite{monetdb}, and an HTAP DBMS: Hyper \cite{hyper}. Unlike these complete DBMSs, LASER is only a storage engine without an SQL query engine. Nevertheless, we include this experiment for a reference comparison against some well-understood DBMSs. We omit a comparison with main-memory-only HTAP systems such as Hyrise \cite{hyrise}, and SAP HANA \cite{HANA}, and with disk-based systems that are not open-source such as SingleStore \cite{singlestore}.

\textbf{Methodology:}
We generate an HTAP workload (\textit{HW}) using queries $Q_1 - Q_5$. To emulate a data lifecycle, we continuously insert new data ($Q_1$) at a steady rate of 10,000 insert operations per second. This ensures that entries continuously move from one level to the next. Along with the inserts, we issue 100 updates per second, i.e., one percent of the insert rate, via $Q_3$, where a randomly chosen column value is updated for a recently inserted key. This update pattern mimics updates and corrections frequently taking place in mixed analytical and transactional processing \cite{stratos2019}. 
Furthermore, we control the access patterns throughout the data lifecycle by selecting $k$, $v$, $v_s$, and $v_e$ for queries $Q_2 - Q_5$ such that the upper levels of the LSM-Tree are mostly accessed by point read operations and wider projections, whereas lower levels are accessed by scan operations and narrower projections. 
This allows us to generate a lifecycle-driven hybrid workload, as described in Section \ref{sec:definitions}.

We use two variants of $Q_2$ for point access of recent data: HW-$Q_{2a}$ and HW-$Q_{2b}$. The $v$ value in each variant is determined by a normal distribution over the time-since-insertion values of the keys. In Figure \ref{fig:exp4-norm98}, we show the two distributions from which $v$ is selected. The mean of the first distribution is 0.98 (typically accessing data from in-memory skiplists, Level-0, or Level-1), and 0.85 for the second distribution (typically accessing data from Level-2 or Level-3); each distribution has a  standard deviation of 0.02. $Q_{2a}$ queries fetch all 30 attributes, whereas $Q_{2b}$ fetches columns 16-30. 

For analytical operations, we use $Q_4$, which accesses columns 21-30 for 5\% of the keys, and $Q_5$, which accesses columns 28-30 for 50\% of the keys. Since our keys are uniformly distributed integer values, these queries access data from all levels, i.e., both recent and historical data.
However, the amount of data scanned at level $i+1$ is a factor $T = 2$ more than that scanned at level $i$. Table \ref{tab:mw} summarizes the properties of these operations. 

We first load 400 million entries, and then execute the workload until another 20 million entries are inserted. Queries HW-$Q_{2a}$ and HW-$Q_{2b}$ are spread uniformly, whereas $Q_4$ and $Q_5$ are executed towards the end. Queries $Q_2$, $Q_4$, and $Q_5$ are issued using four concurrent client threads, whereas a separate client thread is responsible for write operations ($Q_1$ and $Q_3$).

The CG configuration used by LASER for this workload is labelled \textit{D-opt} (Figure \ref{tab:exp-designs}), and was computed as described in Section \ref{sec:best-design-sol}. For comparison, we select five other designs with varying CG sizes. The design with cg\_size=30 corresponds to a pure row-oriented layout (default RocksDB) and the design \textit{rocksdb-col} corresponds to a simulated pure column-oriented layout inside LASER. Since column stores such as MonetDB benefit from contiguous storage of column values, we simulate this in \textit{rocksdb-col} by restricting the LSM-Tree to 2 levels and we set cg\_size=1 (Level-0 absorbs flushed skiplists, and Level-1 stores all the sorted runs with cg\_size=1).  The remaining three designs correspond to CG sizes that match the projections of the operations in the workload \textit{HW}. The design with cg\_size=15 matches $Q_{2b}$, the design with cg\_size=3 matches $Q_5$, and the design with cg\_size=6 is partly suitable for $Q_4$ and $Q_5$. 
We also consider a design we call HTAP-simple, in which 25 percent of the most recent data are stored in a row-oriented layout, and the remaining 75 percent are stored in a pure column-oriented layout.
To test various CG layouts within reasonable time, we opted to have deeper LSM-Trees, therefore, we set the level size ratio (T) to 2.
For all the designs (except \textit{rocksdb-col}), the LSM-Trees have 8 levels with Level-0 in row-format. 
To isolate the impact of the storage layout, we simulate these seven designs within LASER.
For the \textit{HTAP-simple} design, we set cg\_size=30 for the first 6 levels and cg\_size=1 for the last 2 levels (which contain 75 percent of the data).

Additionally, we execute this workload in the following row-store DBMSs: Postgres-9.3, MySQL 5.6 with the MySQL storage engine and MyRocks \cite{myrocks} storage engine, a column-store DBMSs: MonetDB 5 server v11.33.3 \cite{monetdb}, and an HTAP DBMS: Hyper \cite{hyper} (via Hyper API v0.0.14946 \cite{hyperapi}). Hyper API does not allow multiple client threads to simultaneous connect to the same database, therefore we executed the workload via a single client thread.

\begin{table}[]
    \centering
    \small
    \begin{tabular}{|c|c|c|c|}
        \hline
        \textbf{Query} & \textbf{Projection ($k$)} & \textbf{Key ($v$) distribution} & \textbf{Count} \\ \hline
        $Q_1$ & 1-30 & uniform & 10,000/sec\\ \hline
        $Q_{2a}$ & 1-30 & normal,0.98,0.02 & 500,000  \\ \hline
        $Q_{2b}$ & 16-30 & normal, 0.85,0.02 & 500,000 \\ \hline
        $Q_3$ & any 1 of 30 & uniform, 1\% of data & 100/sec \\ \hline
        $Q_4$ & 21-30 & uniform, 5\% of data & 12 \\ \hline
        $Q_5$ & 28-30 & uniform, 50\% of data & 12 \\ \hline
    \end{tabular}
    \caption{Summary of the HTAP workload HW}
    \label{tab:mw}
\end{table}

\textbf{Results:}
Figure \ref{fig:exp4} shows that LASER's optimal design outperforms the other storage layouts when executing the mixed workload described in Table \ref{tab:mw}. Figure \ref{fig:exp4-total-time} shows that LASER took the least total time to execute the complete workload, i.e., queries $Q_1$ to $Q_5$. Designs with cg\_size=2, MySQL, MyRocks, MonetDB, and Hyper did not finish within our time-limit-exceeded (TLE) window of 24 hours. Therefore, we instead report their average latencies in Figure \ref{fig:exp4-read} and \ref{fig:exp4-scan}.

Figure \ref{fig:exp4-insert-throughput} compares the insert throughput during the load phase. LSM-Trees are known to stall inserts as compaction tasks queue up because compaction involves slow disk IOs whereas inserts happen in memory \cite{writeStalls}. Thus, the insert throughput of each design is dependent on the amount of compaction. From Equation \ref{eq:insert}, the amount of compaction (i.e., write amplification) depends on the number of CGs and the number of levels, $L$. Design \textit{rocksdb-col} has the smallest compaction size since it has only two levels, and thus the highest throughput. However, \textit{rocksdb-col} simulates a pure column-oriented layout with only 2 levels, whereas in practice LSM-Trees have 8 or more levels \cite{rocksdbTuning}. Amongst the other designs (with 8 levels), \textit{rocksdb} has the highest throughput because it has the fewest column groups. LASER's throughput is 25 percent slower than \textit{rocksdb}'s since it has multiple column groups. LASER's design selection (Section \ref{sec:best-design-sol}) trades off the insert throughput to achieve better query latency, as we will see in Figure \ref{fig:exp4-read} and \ref{fig:exp4-scan}, optimizing for the overall workload time. If high insert throughput is more critical than query performance then the cost of inserts in Equation \ref{eq:cost} can be multiplied by some weight, which would force LASER to select a more write-optimized design. Postgres, MySQL, MyRocks, MonetDB, and Hyper performed significantly worse, to the extent that loading data via a stream of INSERT statements was impractical. To finish the data load within reasonable time, we loaded directly from the files, and we exclude these systems from the throughput comparison.

In Figure \ref{fig:exp4-read}, LASER's design has either the lowest latency or is close to the lowest latency across different designs.
MySQL, a row oriented DBMS, has $Q_2$ latency very close to LASER.
MonetDB has the highest latency, orders of magnitude slower than LASER. Insert and update latencies across all the LSM-Tree designs (including LASER's) are the same because they all append the data to an in-memory skiplist, which is not impacted by the layout of the disk levels. Hyper, which supports HTAP workloads, like LASER, performs significantly worse for $Q_2$ and $Q_3$. This is because it does not consider the data lifecycle while deciding the storage layout.  Instead, Hyper stores all the data in columnar layout and only varies the compression scheme for hot and cold data \cite{datablocks}. In Figure \ref{fig:exp4-scan}, LASER's latency is close to the latency of the design best suitable for the query (e.g., cg\_size=3 is suitable for $Q_5$ and cg\_size=15 is suitable for $Q_4$).
For $Q_5$, MonetDB and Hyper perform 5x better than LASER because they store all the data in contiguous columns, which is suitable for aggregation queries. However, MonetDB performs 20x worse than LASER for $Q_4$. Hyper performs the best for $Q_4$ partly due to the columnar layout and partly because of a single client workload. We suspect Hyper's latency for $Q_4$ will increase with multiple parallel clients due to disk throttling. Postgres was the best performing row-store DBMS, with similar latency as LASER for $Q_4$, but it was 2x slower for $Q_5$.

\vspace{-3pt}
\subsection{Robustness to Workload Shifts}
\label{sec:stress-testing}
\textbf{Goal:}
We now measure the performance impact when the tested workload deviates from the representative workload.

\textbf{Methodology:}
We consider two types of shifts from the workload \textit{HW} used in Section \ref{sec:exp4}: 1) vertical shift in read access patterns, and 2) horizontal shift in scan projections. For the vertical shift, we offset the mean of the normal distribution used to generate keys for $Q_{2a}$ and $Q_{2b}$ in\textit{HW}. For example, an offset of 0.1 results in a mean of 0.88 for the distribution generating keys for $Q_{2a}$ and a mean of 0.75 for $Q_{2b}$. 
For the horizontal shift, we offset the projection of $Q_5$ in \textit{HW} to the left by some amount. For example, offsetting the projection by 2 results in projecting columns 26-28, and a shift of 4 results in columns 24-26, and so on. 

\begin{figure}[t]
\begin{subfigure}{0.25\textwidth}
  \centering
  \includegraphics[width=4.6cm,height=2.5cm]{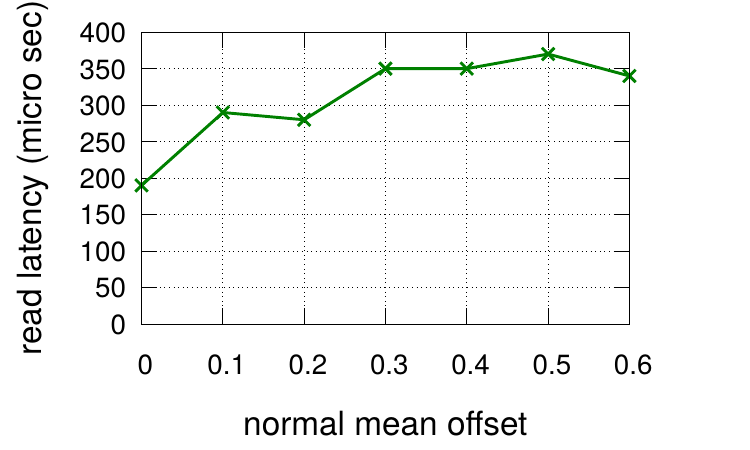}  
  \caption{{Change in read latency \\ with shifting read pattern}}
  \label{fig:stress-read}
\end{subfigure}%
~
\begin{subfigure}{0.24\textwidth}
  \centering
  \includegraphics[width=4.6cm,height=2.5cm]{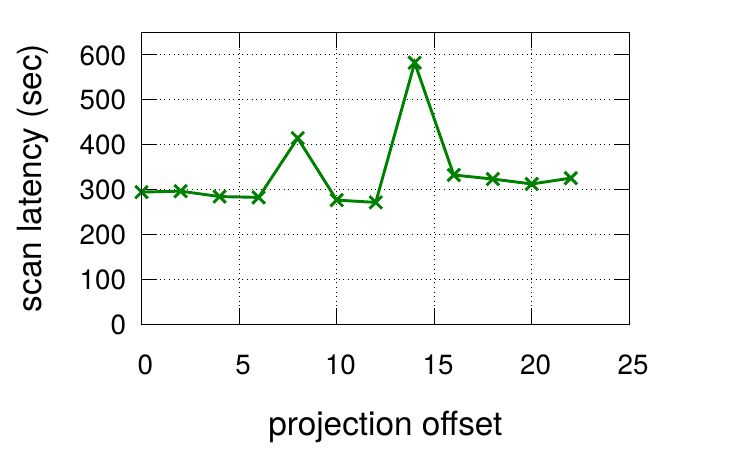}  
  \caption{{Change in scan latency \\ with shifting projections}}
  \label{fig:stress-scan}
\end{subfigure}
\caption{{LASER is robust to minor workload shifts}}
\label{fig:exp4-stress}
\end{figure}

\textbf{Results:}
Figure \ref{fig:stress-read} shows the results of vertical shifts in the read pattern.
The latency of read operations initially increases with the shift but then stays constant. This is because top levels are small, so the CG layout changes with the offset, but eventually when the offset fetches keys from the last few levels (which are larger), the CG layout does not change with the offset.

In Figure \ref{fig:stress-scan}, we examine horizontal shifts in the scan projection.
Scan latency gets up to 2x worse if the projections are misaligned and fetch data from a few wide CGs. For example, when the offset is 14, the projection is 14-16, which spans two large CGs, <1-15> and <16-20>, in levels 6 and 7.
However, if the CGs are small, or only a single CG is required by the projection, the impact of misalignment is smaller. As expected, from Figures \ref{fig:stress-read} and \ref{fig:stress-scan}, we conclude that the performance of read and scan operations deteriorates if the mismatch between the actual workload and the representative workload is significant. 
In these cases, LASER should be re-configured. We consider a self-configuring Real-Time LSM-Tree as a direction for future work as part of \textit{online tuning of physical design} \cite{bruno}.

\section{Related Work}
\label{sec:related-work}

\textbf{Adoption of LSM-Trees:}
LSM-Trees are used in many RDBMSs, key-value stores and NoSQL systems \cite{oneil}. 
Other applications include the log-structured history access method (LHAM) \cite{lham}, which supports temporal workloads by attaching timestamp ranges to sorted runs and pruning irrelevant sorted runs at query time. Furthermore, LSM-trie \cite{lsm-trie} is a hash index for key-value pairs where the metadata, such as index pages, cannot be fully cached. 
Finally, the LSM-based tuple compaction framework in AsterixDB \cite{tupleCompaction} leverages LSM lifecycle events (flushing and compaction) to extract and infer schemas for semi-structured data. Similarly, in LASER, we exploited LSM-Tree properties, such as data propagation through the levels over time and compaction.

\textbf{Improvements of LSM-Trees:}
Recent works have optimized various components of LSM-Trees such as allocating space for Bloom filters \cite{monkey}, tuning the compaction strategy \cite{dostoevsky}, and compaction scheduling to mitigate write stalls
\cite{writeStalls}.
Many of these recent improvements are orthogonal to the design of LASER. 

\textbf{HTAP systems and storage engines:}
An early approach, fractured mirrors, maintained one copy of the data in row-major layout and another copy in column-major layout \cite{fracturedMirrors}. This has been adopted by Oracle and IBM to support columnar layout as an add-on. Although these systems achieve better OLAP performance than a pure row-store, the cost of synchronizing the two replicas is high. HYRISE \cite{hyrise} partitions tables into column groups based on how columns are co-accessed by queries. Systems such as SAP HANA \cite{HANA}, SingleStore \cite{singlestore}, and IBM Wildfire \cite{wildfire} split the storage into OLTP friendly and OLAP friendly components. Data are ingested by the OLTP friendly component, which is write-optimized and uses a row-major layout, and are eventually moved to the OLAP friendly component, which is read-optimized and uses a column-major layout. Peloton (now NoisePage) \cite{pavlo, noisepage} generalizes this idea by partitioning the data into multiple components called tiles,
with different column group layouts.
In this work, we described these systems as having a \textit{lifecycle-aware} data layout, and we showed that LSM-Trees are a natural fit for a lifecycle-aware key-value storage engine.


\section{Conclusions}
\label{sec:conclusion}
We showed that Log-Structured Merge Trees (LSMs) can be used to design a lifecycle-aware storage engine for HTAP systems.  To do so, we proposed the idea of a Real-Time LSM-Tree, in which different levels can store the data in different formats, ranging from purely row-oriented to purely column-oriented.  We presented a design advisor to select an appropriate Real-Time LSM-Tree design given a representative workload, and we implemented a proof-of-concept prototype, called LASER, on top of the RocksDB key-value store.  


\section{Acknowledgments}
We would like to thank Andy Yu for providing the experimental results for Hyper DBMS.
\bibliographystyle{ACM-Reference-Format}
\bibliography{paper}
\end{document}